\documentclass[11pt,preprint]{aastex}
\begin{document}

\slugcomment{accepted to the {\it Astrophysical Journal}}

\title{The Initial Mass Function of Low-Mass Stars and Brown Dwarfs in Young 
Clusters\altaffilmark{1,2}}

\author{K. L. Luhman}
\affil{Harvard-Smithsonian Center for Astrophysics, 60 Garden St., 
Cambridge, MA 02138}

\email{kluhman@cfa.harvard.edu}

\author{G. H. Rieke, Erick T. Young, Angela S. Cotera,
H. Chen, Marcia J. Rieke, Glenn Schneider, and Rodger I. Thompson}
\affil{Steward Observatory, The University of Arizona, 933 North Cherry Avenue, 
Tucson, AZ 85721}

\email{grieke, eyoung, cotera, hchen, mrieke, gschneider,
rthompson@as.arizona.edu}

\altaffiltext{1}{Based on observations made with the Multiple Mirror Telescope 
operated by the Smithsonian Astrophysical Observatory and the University of 
Arizona.}

\altaffiltext{2}{Based on observations made with the NASA/ESA Hubble Space
Telescope, obtained at the Space Telescope Science Institute, which is operated
by the Association of Universities for Research in Astronomy, Inc., under NASA
contract NAS 5-26555. These observations are associated with proposal ID 7217.}

\begin{abstract}

We have obtained images of the Trapezium Cluster
($140\arcsec\times140\arcsec$; 0.3~pc~$\times$~0.3~pc) with
the Hubble Space Telescope Near-Infrared Camera and Multi-Object Spectrometer 
(NICMOS). Combining these data with new ground-based $K$-band spectra ($R=800$) 
and existing spectral types and photometry, we have constructed an 
H-R diagram and used it and other arguments to infer masses and ages. 
To allow comparison with the results of our previous studies of 
IC~348 and $\rho$~Oph, we first use the models of 
D'Antona \& Mazzitelli. With these models, the distributions of ages of 
comparable samples of stars in the Trapezium, $\rho$~Oph, and IC~348 indicate 
median ages of $\sim0.4$~Myr for the first two regions and $\sim1$-2~Myr for 
the latter. The low-mass IMFs in these sites of clustered star formation are 
similar over a wide range of stellar densities 
($\rho$~Oph, $n=0.2$-$1\times10^3$~pc$^{-3}$; IC~348, $n=1\times10^3$~pc$^{-3}$;
Trapezium, $n=1$-$5\times10^4$~pc$^{-3}$) and other environmental conditions 
(e.g., presence or absence of OB stars). With current data, we cannot rule out 
modest variations in the substellar mass functions among these clusters.
We then make the best estimate of the true form of the IMF in the Trapezium 
by using the evolutionary models of Baraffe et al.\ and an empirically 
adjusted temperature scale 
and compare this mass 
function to recent results for the Pleiades and the field. All of these data 
are consistent with an IMF that is flat or rises slowly from the substellar 
regime to about 0.6~$M_{\odot}$, and then rolls over into a power law that 
continues from about 1~$M_{\odot}$ to higher masses with a slope similar to 
or somewhat larger than the Salpeter value of 1.35. For the Trapezium, this 
behavior holds from our completeness limit of $\sim0.02$~$M_{\odot}$ and 
probably, after a modest completeness correction, even from 
0.01-0.02~$M_{\odot}$. These data include $\sim50$ likely brown dwarfs.
We test the predictions of theories of the IMF against 1) the shape of the 
IMF, which is not log-normal, in clusters and the field, 2) the similarity 
of the IMFs among young clusters, 3) the lowest mass 
observed for brown dwarfs, and 4) the suggested connection between the 
stellar IMF and the mass function of pre-stellar clumps. 
In particular, most models do not predict the formation of the moderately 
large numbers of isolated objects down to 0.01~$M_{\odot}$ that we find in 
the Trapezium. 

\end{abstract}


\section{Introduction}

What is the true form of the stellar initial mass function (IMF)? Is it 
universal, or does it depend on the properties of the natal molecular cloud 
or of the embedded, young stellar population?  In particular, the turnover mass and the minimum 
mass of the IMF and their behavior with various star forming conditions
can offer vital insights into the physical processes that 
regulate the formation of stars and brown dwarfs (Elmegreen 1999b).

Numerous techniques have been used to measure the IMF
(see reviews by Scalo 1998; Elmegreen 1999b).  Stellar open
clusters have played an important role, but difficulties arise in 
determining cluster membership and completeness at low masses, and in 
properly accounting for dynamical evolution and mass segregation.
These problems can be alleviated in the youngest clusters ($\leq10$~Myr)
associated with star forming regions. The compact nature and thick molecular 
cloud of a star-forming cluster can greatly reduce contamination by foreground 
and background stars. Newborn substellar objects are quite luminous 
and should have the same spatial distribution as the stars since
these regions are too young to have undergone significant 
dynamical evolution. Furthermore, both the initial conditions of 
star formation and the resulting mass function are directly observable in the 
youngest clusters. Stellar populations in clusters are also relevant for 
comparisons to the field and other regions since it is likely that clusters 
represent the dominant mode of star formation in the Galaxy (Lada, Strom, \& 
Myers 1993).

The Orion Nebula Cluster centered on the Trapezium OB stars is the
richest of any nearby cluster and has been studied extensively through proper 
motions (Jones \& Walker 1988), optical images from the ground 
(Herbig \& Terndrup 1986) and space (Prosser et al.\ 1994), wide-field infrared 
(IR) images (Ali \& DePoy 1995), and high-resolution ground-based IR images 
(McCaughrean \& Stauffer 1994, hereafter MS; 
Petr et al.\ 1998; Simon, Close, \& Beck 1999). Hillenbrand (1997) 
combined new optical spectroscopy and photometry with previous data from the 
literature for more than 1000 stars within $18\arcmin$ (2.5~pc) of the 
Trapezium OB stars and constructed a Hertzsprung-Russell (H-R) diagram for 
the Orion cluster. With the theoretical evolutionary models of D'Antona \& 
Mazzitelli (1994) (hereafter DM94), she inferred an average age of $<1$~Myr 
and an IMF that peaked at 0.2~$M_{\odot}$ and fell rapidly to lower masses.
Hillenbrand \& Carpenter (2000) have recently extended this work to substellar 
masses through deep ground-based $H$ and $K$ imaging of the 
$5\farcm1\times5\farcm1$ area surrounding the Trapezium OB stars.  

Within the central $\sim5$~arcmin$^2$ of the Orion Nebula Cluster lies the
Trapezium Cluster, where stellar densities reach a peak of
$\sim5\times10^4$~pc$^{-3}$ (MS). Several characteristics make the Trapezium 
Cluster a unique region for a study of the IMF. The cluster is rich (300 stars) 
and nearby (450~pc) and a majority of its members have minimal extinction 
($A_V<5$) because of the cluster's location on the front of the molecular 
cloud and within the cavity created by the O star $\theta^1$~Ori~C.
In addition, the obscuration of the molecular cloud and the compactness
of the cluster minimize contamination from background and foreground stars.
Because of the abrupt change in reddening from the Trapezium members to 
the field stars behind the cloud, virtually all sources with $A_V<10$ should be 
cluster members, a crucial property in reliably identifying the substellar 
population. The special viewing geometry for the Trapezium Cluster overcomes 
many of the common limitations in correcting cluster measurements for
the effects of obscuration.  
Consequently, it should be possible to construct an
IMF from well below the hydrogen burning mass limit to about 50~$M_{\odot}$.

In \S~\ref{sec:trap} of this paper, we describe a new study of the Trapezium 
Cluster.  Previous observations of the low-mass population in the Trapezium 
have been hindered by crowding and bright nebulosity. To overcome these 
obstacles, we have obtained sensitive ($H\sim17$) high-resolution ($0\farcs2$) 
images of the Trapezium Cluster ($140\arcsec\times140\arcsec$) with the 
Near-Infrared Camera and Multi-Object Spectrometer (NICMOS) aboard 
the {\it Hubble Space Telescope (HST)}. We have also
measured $K$-band spectra for $\sim100$ sources in this region. We use these 
measurements and data from the literature (MS; Prosser et al.\ 1994; 
Hillenbrand 1997) to construct an H-R diagram for the cluster and infer 
individual masses and ages from theoretical evolutionary models (D'Antona \& 
Mazzitelli 1997, hereafter DM97; Baraffe et al.\ 1998, hereafter B98). 
The resulting star formation history is used to estimate
masses for the faint sources that lack spectral types, which are combined
with the masses of stars on the H-R diagram to produce a cluster IMF
that reaches down to 0.01~$M_{\odot}$.

In \S~\ref{sec:compare}, we compare the Trapezium IMF
with the similarly derived mass functions for the star-forming clusters 
IC~348 (Luhman et al.\ 1998, hereafter LRLL; Luhman 1999) and $\rho$~Oph 
(Luhman \& Rieke 1999, hereafter LR99), showing them
all to be similar. Interpretation of young cluster observations is limited by
the accuracy of the theoretical models for the evolution
of young stars and brown dwarfs. It is now possible to mitigate
this problem by testing evolutionary tracks and temperature
scales against young multiple star systems that contain
coeval stars of differing mass. If we use a combination
of tracks (B98) and temperature scale that is closely 
consistent with such observational tests (see Luhman 1999), the IMF for the 
Trapezium and the other clusters is similar to 
recently determined mass functions for the Pleiades and M35 open clusters and 
the field. The shape of the low-mass IMF, its approximate invariance
across two orders of magnitude in stellar density, and the presence
of moderately large numbers of very low-mass brown dwarfs are used to test the 
predictions of various theoretical models for the origin of the IMF.

\section{The Trapezium Cluster}
\label{sec:trap}

\subsection{Observations and Data Analysis}
\label{sec:obs}

\subsubsection{$K$-band Spectroscopy}

We performed $K$-band spectroscopy on sources in the Trapezium Cluster 
using the near-IR long-slit spectrometer FSpec (Williams et al.\ 1993) 
at the Multiple Mirror Telescope on Mount Hopkins on the nights of 
1995 November 8 and 10, 1995 December 2 and 3, and 1996 February 5, 6, 9, and 
10.  The wavelength coverage was 2.0 to 2.4~\micron\ with a 
two-pixel resolution of $R=\lambda/\Delta\lambda=800$. The observations 
and data reduction procedures were identical to those described by LRLL.
From the $K$-band photometry of the central square arcminute of the Trapezium
by MS, we selected for spectroscopy 64 of 
the 77 sources with $K<12$ and 8 sources that were somewhat fainter. 
In addition, we observed 29 stars with $K<11$ appearing in images 
outside of the region of MS (M. McCaughrean, private communication),
which included several embedded sources in the BN/KL nebula 
(Becklin \& Neugebauer 1967; Kleinmann \& Low 1967). 
On 1996 December 28 we used a new grating that provided $R=1200$ for 
follow-up observations of source~n from Lonsdale et al.\ (1982) 
and the BN object (stars 50 and 44 in Table~1).

\subsubsection{NICMOS Photometry}

On 1998 January 30 we observed the Trapezium Cluster with 
camera 3 of NICMOS (NIC3) on {\it HST}.
At a plate scale of $0\farcs201\pm0\farcs001$, NIC3 provides a field
of view of $51\farcs2\times51\farcs2$. Nine contiguous pointings were imaged
in a $3\times3$ dither pattern where the corners of the total 
$140\arcsec\times140\arcsec$ field have coordinates of $(\alpha,\delta) (2000)=
(5^{\rm h}35^{\rm m}12\fs11$, $-5\arcdeg21\arcmin48\farcs3$), 
($5^{\rm h}35^{\rm m}11\fs87$, $-5\arcdeg24\arcmin09\farcs8$), 
($5^{\rm h}35^{\rm m}21\fs61$, $-5\arcdeg21\arcmin51\farcs6$), and
($5^{\rm h}35^{\rm m}21\fs37$, $-5\arcdeg24\arcmin13\farcs1$).
Images were obtained through the F110W (0.8-1.4~\micron) and F160W 
(1.4-1.8~\micron) filters with integration times of 96 and 80~s, respectively.
Dark frames were taken during the observations.
Dark subtraction and linearity corrections were performed with 
the NICRED data reduction package (McLeod 1997).

Because of the number and uneven distribution of bright stars within
the field of view, the background level varied considerably among the 
quadrants of the array and among the dithered frames. 
The offsets between quadrants were interactively determined by minimizing the 
median of the differences between border pixels.  The upper right quadrant
was always assumed to be correct, then the two adjacent quadrants were
offset to match this quadrant. Finally, the last quadrant was
adjusted to minimize the difference with the two adjacent quadrants.
In constructing the mosaic images in Figs.~\ref{fig:f110} and \ref{fig:f160},
offsets between the dithered frames were measured in a similar fashion.
The quadrant offsets were applied prior to flat-fielding.

Stellar coordinates and photometry were measured from the NICMOS images
with the package APPHOT within the IRAF environment.
Initial identifications of sources with DAOFIND were checked 
through visual inspection. Given the high spatial resolution 
of the data, knots of nebulosity were easily rejected. 
A few sources accepted as stars have slightly extended profiles
(45, 103, 215 in Table~1), possibly due to circumstellar material.  
Because the point-spread function (PSF) of {\it HST} is undersampled by NIC3, 
it can be difficult to distinguish a faint star from a cosmic ray hit.  
However, nearly all objects that were identified by 
DAOFIND and through visual inspection were detected in at least 
two bands in the NICMOS and ground-based photometry (including unpublished
$K$-band measurements of M. McCaughrean). An exception is source 65, which
was quite faint and detected only in F160W. Several faint $K$-band sources 
of MS could not be measured with NICMOS due to their close proximity to 
the OB stars.

Aperture photometry was extracted for all sources with the task PHOT using
a radius of two pixels. The background level was measured
in an annulus around each source and subtracted from 
the photometry. Because of the structure of the
nebulosity, the background was measured as close as possible to each star
by using an annulus one pixel wide.
The inner radii of these annuli ranged from three to six
pixels, where the larger annuli were required to sample the 
background emission properly around brighter stars.

The data were calibrated assuming $2.873\times10^{-6}$ and
$2.776\times10^{-6}$~Jy~ADU$^{-1}$~sec$^{-1}$ and zero-magnitude fluxes of 
1775 and 1083~Jy on the Vega system respectively for F110W and F160W.
To apply this calibration, it is necessary to correct our
small-aperture photometry to total signal. From bright stars in
our images, we measured the aperture corrections from a radius
of 2 to a radius of 7.5~pixels to be 0.125 and 0.150~mag for F110W
and F160W, respectively. Additional corrections from 7.5 to 22.5~pixels
were estimated from simulations of the PSF to be 0.030
and 0.046~mag for F110W and F160W.
Stars brighter than m$_{110}\sim10$ and m$_{160}\sim9$ are saturated. 
Due to the variations in the nebulosity, the detection limits 
are not constant across the field, but are typically m$_{110}\sim18$ 
and m$_{160}\sim17$.

Except at these faintest limits, the photometric
uncertainties are dominated by the undersampling of the {\it HST} PSF 
by the NIC3 detector. Because of the overlap among the images in the
$3\times3$ grid of pointings, we have more than one measurement 
for a large number of objects. The differences in the separate measurements
were 0.2~mag or less for most sources, hence 
an approximate error in the photometry is no more than $\pm0.2$~mag. The 
m$_{110}-{\rm m}_{160}$ colors, on the other hand, showed a smaller
scatter. The separate pointings of {\it HST} at F110W and F160W may have been 
precise enough to produce similar samplings of the PSF, resulting in 
relatively accurate colors. The average of the available measurements
for a given star is taken as the final photometry listed in Table~1.

To derive coordinates for the NICMOS sources, we used the stars in the 
overlapping regions to compute offsets among the nine F160W frames
and to place the stars on the same pixel coordinate 
system. A plate solution was measured with coordinates of 
non-saturated sources detected in $K$-band images of the NICMOS 
field (M. McCaughrean, private communication). For stars that are saturated
in the NICMOS frames, we adopted the $K$-band coordinates. The $K$-band data of 
McCaughrean do not provide precise coordinates for sources
127, 153, 154, and 191 and these stars are too close to bright 
stars to be measured in the NICMOS frames. We used the offsets of these stars
from their bright neighbors as provided by MS to place them on the  
coordinate system (the declinations are the same and
the right ascensions are 0.013~sec greater in
the coordinate system of MS compared with that of McCaughrean). 

Table~1 lists all known optical and near-IR point sources within the 
NICMOS field ($140\arcsec\times140\arcsec$) towards the Trapezium.
Several close pairs in the {\it HST} images of
Prosser et al.\ (1994) are unresolved in the NICMOS and ground-based 
data. These pairs are treated as one object and in \S~\ref{sec:colors} the 
luminosities and reddenings are calculated from the combined photometry 
for the system. There is one fairly bright object ($K=14$)
that was observed by MS but not detected in the NICMOS images. It is
probably very red or a knot of nebulosity that was unresolved in the 
ground-based data. Faint companions in the NICMOS images that are not 
detected in the $K$-band images include objects 67, 70, 218, 255, and 303.
Because TCC075 and TCC077 are only partially resolved in the NICMOS data,
photometry was extracted from the combined system. The $K$-band measurements
of MS for these two objects were combined to produce the value listed in 
Table~1. The star ID459 from the compilation of Hillenbrand (1997) is reported 
to have $I_C=12.5$, but has no measurement at $V$ and is not detected 
in any of the IR data. This object is probably a ghost in the images of
Jones \& Walker (1988).  In the NICMOS images, TCC009 is nebulous and does 
not appear to be stellar. ID459 and TCC009 are excluded from Table~1. 

\subsection{Individual Source Characteristics}

\subsubsection{A New CO Emission Source}

We have detected second overtone CO emission towards object 50, otherwise
known as source~n (Lonsdale et al.\ 1982).  Spectra for this source and the 
BN object are shown in Figure~\ref{fig:spec}.  An $H$-band spectrum of source~n
is featureless at a signal-to-noise of 25.
Source~n and the BN object are the reddest objects with detections
at both NICMOS bands, with colors of m$_{110}-{\rm m}_{160}=4.25$
and 5.44 respectively. Given the fairly high luminosity for source~n implied 
by its $K$ magnitude and very red colors, the CO emission probably arises from 
an inversion layer in an irradiated disk around a luminous, hot central star 
(Calvet et al.\ 1991; Biscaya Holzbach et al.\ 1999). Source~n had received 
little attention until Menten \& Reid (1995) observed a double radio source 
centered on it. Since CO emission usually is accompanied by outflows, our 
observations agree with and complement their proposal that source~n drives
a maser outflow and could contribute significantly
to the energetics of this part of the BN/KL nebula.

\subsubsection{Spectral Types and Extinctions}
\label{sec:sptypes}

We determined spectral types for the Trapezium Cluster
in a manner similar to our previous studies of other clusters (LR98,
LRLL, and LR99); therefore, we have placed the detailed description
of this process in the appendix. For similar reasons, we have
used the same appendix for a discussion of the extinction determination
and correction. We define the Trapezium spectroscopic sample as all objects in 
Table~1 that have spectral types, either IR types from this work and/or 
previously published optical types.

\subsubsection{Surface Gravities}
\label{sec:gravities}

Because the ratio of Na and Ca to the CO band heads changes significantly 
with luminosity class (Kleinmann \& Hall 1986), $K$-band spectra can be used 
to examine qualitatively the surface gravities of young stars.
In previous studies by Greene \& Meyer (1995) and Greene \& Lada (1997),
the relative strengths of the atomic lines and the CO band heads in young
stars have appeared intermediate between the values for dwarfs and 
giants. This effect is largely due to the deepening of CO with lower 
surface gravity at a given spectral class, although the simultaneous 
weakening of Na in mid-to-late M stars is also a significant contributor (LRLL).

In Figure~\ref{fig:width}, the first band head of CO is plotted versus the 
sum of Na and Ca for the Trapezium, the $5\arcmin\times5\arcmin$ core of 
IC~348 (LRLL), and the cloud core of $\rho$~Oph (LR99).
The solid lines represent fits to the measured equivalent widths of standard 
dwarfs and giants in LR98. 
The dwarf locus is shown for spectral types of M4V and earlier. At later
types, the CO continues to strengthen while the atomic lines weaken (see LR98).
The late M stars in the IR spectroscopic sample are faint and have low 
signal-to-noise data, hence the weak atomic lines cannot be measured 
accurately and we omit objects later than M4 in this analysis. As illustrated 
in Figure~\ref{fig:width}, the $K$-band spectra indicate lower average surface 
gravities from IC~348 to $\rho$~Oph to the Trapezium, with a larger
spread in gravities in IC~348. Since surface gravities should be
lowest at the earliest stages of stellar evolution,
we compare these results to the distributions of ages of these clusters 
implied by the DM97 evolutionary models in \S~\ref{sec:ages}.
IC~348 is clearly older than the Trapezium
and $\rho$~Oph from both the inferred ages and the surface gravity diagnostics.
The IR spectroscopic sample for the Trapezium in
Figure~\ref{fig:width} has the same distribution of ages as the entire 
spectroscopic sample (optical$+$IR) discussed in \S~\ref{sec:ages}. This 
distribution is younger than that in the $\rho$~Oph cloud core, in agreement
with the slightly lower surface gravities suggested for the Trapezium in
Figure~\ref{fig:width}. However, when we account for selection
biases in \S~\ref{sec:ages}, we find that the Trapezium and the cloud core of
$\rho$~Oph have similar ages.

\subsubsection{Effective Temperatures and Bolometric Luminosities}
\label{sec:teff}

Reddenings, effective temperatures, and luminosities
for the six OB stars in Table~1 are taken from Hillenbrand (1997).
For the remainder of the sample, we used the reddening
corrections discussed in the appendix and the spectral classifications and
photometry to estimate temperatures and luminosities.

Spectral types of M0 and earlier
are converted to effective temperatures with the dwarf temperature scale 
of Schmidt-Kaler (1982). LR98 considered the various M dwarf temperature scales
and concluded that the conversion of Leggett et al.\ (1996) agreed with the
data on the two eclipsing binaries, CM~Dra and YY~Gem.  Luhman (1999) recently 
used the components of the young quadruple system GG~Tau (White et al.\ 1999)
and the locus of stars in IC~348 as empirical isochrones to test combinations 
of theoretical evolutionary models
and temperature scales for very young objects. Although
there is some discrepancy near the hydrogen burning limit, 
the dwarf temperature scale was compatible with the models of DM97 at higher
and lower masses. Therefore, we will use the dwarf scale when our data are 
interpreted with the DM97 models, just as in our studies of
IC~348 and $\rho$~Oph. As a result, we will be able to compare 
the IMFs confidently among these clusters. The temperatures listed in Table~1 
correspond to the average of the adopted spectral type ranges under the 
dwarf scale. The temperature scale that produces agreement between the 
model isochrones of B98 and the data for 
IC~348 and GG Tau is intermediate between the scales for M dwarfs and giants 
(see Figure 7 of Luhman 1999). Therefore, we will use this temperature scale 
when calculating the IMF with the models of B98 in \S~\ref{sec:imf}.

The photometric bands $R$ through $H$ are the least susceptible to
contamination from short or long wavelength excess emission (e.g., 
Meyer, Calvet, \& Hillenbrand 1997). For the Trapezium
data, the bolometric luminosities are measured from m$_{160}$ except
for a few cases discussed in the appendix where m$_{160}$
was not available and we used $I_C$.
With the models of Allard et al.\ (1998), we estimate
m$_{160}-H=0.1$, 0.04, and 0.02 at effective
temperatures of 3000, 4000, and 6000~K. A color of m$_{160}-H=0.05$ is adequate
for converting all BC$_H$ to BC$_{160}$, where the dwarf bolometric corrections
are the ones used by Luhman (1999). We arrive at the bolometric luminosities 
in Table~1 by combining the bolometric corrections, dereddened m$_{160}$ 
or $I_C$, and a distance modulus of 8.27 (Genzel \& Stutzki 1989). 
The typical uncertainties in the luminosities are $\pm0.15$-0.2~dex in 
log~$L_{\rm bol}$.  These rather large errors are a result of reddening 
uncertainties due to variability between the observations at $I_C$ and 
m$_{160}$, and to the undersampling of the NICMOS PSF, which affects all our 
NICMOS photometry.  

\subsection{The Trapezium Stellar Population}

\subsubsection{Cluster Membership}
\label{sec:member}

Given the very small area of sky covered by the Trapezium Cluster
and the distance of Orion out of the Galactic plane, we expect 
very little contamination from field stars in the foreground.
Stars 125 and 129 were rejected as nonmembers by Hillenbrand (1997) because
proper motions were detected. The colors of these stars are redder than 
expected for foreground stars and one object appears to have enhanced CO 
absorption, supporting a pre-main-sequence nature. Nevertheless, we omit these 
stars from the remainder of our analysis.  The C$^{18}$O maps of 
Goldsmith, Bergin, \& Lis (1997) indicate a large amount of extinction
throughout the NICMOS field, with a gradient from 
east ($A_V\sim30$) to west ($A_V\sim100$). 
Because of this heavily obscuring molecular cloud behind the Trapezium, no
background stars should appear except at very large reddenings.
We exclude background stars by including in our sample 
only stars with reddenings of $A_H\leq1.4$, equivalent to $A_V\leq8$.
As seen in Figure~\ref{fig:hjh}, the distribution of colors 
remains constant as a function of magnitude within the low reddening sample, 
implying that the fainter sources are indeed an extension of the Trapezium 
population to substellar masses and not a contaminating population
of heavily reddened background stars. The unique viewing geometry of 
the Trapezium Cluster has allowed us to detect this rather large population
of brown dwarfs and thus measure the IMF to very low masses. 

\subsubsection{The H-R Diagram}
\label{sec:hr}

To allow comparison of the Trapezium IMF to the previous measurements for 
IC~348 and $\rho$~Oph, we first use the models of DM97 to infer masses
from the source temperatures and luminosities.  We then use the models of B98 
to obtain the most accurate possible mass estimates to constrain the 
true form of the Trapezium IMF.
B98 advise caution in the use of their models for very young clusters 
because they use atmospheres limited to surface 
gravities of log~$g\geq3.5$, which corresponds to ages of $\gtrsim1$~Myr.
Nevertheless, as discussed by Luhman (1999) and White et al.\ (1999), the
B98 models produce better agreement than any others
with the few dynamical mass estimates for young stars
and thus we adopt them as the most accurate available. 

Because the B98 models do not include
stars above 1~$M_{\odot}$, the B98 IMF is constructed in a somewhat different 
fashion than with DM97. As shown in Figure~\ref{fig:hr}, the mass tracks 
of DM97 and B98 differ significantly for very young stars near 1~$M_{\odot}$. 
However, at higher masses and warmer temperatures, uncertainties in the
treatment of convection are less important and the various models (e.g., DM97; 
Bernasconi 1996; Swenson, private communication) agree fairly well at 
2-3~$M_{\odot}$. Thus, for the B98 IMF we use the models of DM97 to infer 
masses at $\geq2$~$M_{\odot}$. To include the stars that fall above the 
1~$M_{\odot}$ track of B98 and below the 2~$M_{\odot}$ track of DM97, we place
these objects in one mass bin extending from log~$M=-0.05$ to 0.35 
(0.89-2.2~$M_{\odot}$). This bin size is then used for the remainder of the
B98 IMF as well, with a division by two to produce the same normalization as
in the DM97 IMF. 

The temperatures corresponding to the average of the adopted spectral
type ranges and the luminosities for the Trapezium spectroscopic sample
are plotted with the evolutionary models in the H-R diagrams in 
Figure~\ref{fig:hr}.  As discussed in \S~\ref{sec:teff}, for the M stars
different temperature scales are used with the tracks of DM97 and B98.
The highly reddened stars in Figure~\ref{fig:hr} are 
generally the youngest, particularly the ones embedded near the BN object. 

Because of the large uncertainties in 
spectral types for some sources, for each star in the Trapezium sample
we assigned 10 spectral types evenly spaced across the adopted range
of types and calculated the resulting luminosities and reddenings. The 
spectral type uncertainties probably do not follow a Gaussian function, thus
we simply assume a uniform distribution. Each of the 10 masses and ages 
inferred for a star is given a weighting of 0.1 and added to the
IMF and the distribution of ages.

\subsubsection{The Distribution of Ages}
\label{sec:ages}

The distribution of ages implied by DM97 for the
Trapezium ($A_H\leq1.4$) is shown in Figure~\ref{fig:ages1}.
Although the youngest isochrone shown in Figure~\ref{fig:hr} is for 0.3~Myr, 
most of the tracks of DM97 do include ages close to 0.01~Myr. 
The few stars falling above this isochrone were placed in the youngest age bin. 
The OB stars have been excluded in Figure~\ref{fig:ages1}. 
The spectroscopic sample in the 
Trapezium is complete only for types earlier than M1 with a bias toward
young objects for later types. We have omitted Trapezium sources later than 
M0 to derive a distribution that is representative of the cluster.

Although the absolute ages implied by evolutionary models must be interpreted
with caution at such early stages, it is instructive
to compare the distributions of ages from cluster to cluster.
The models can indicate different ages as a function of mass in
the same cluster, hence we must consider similar ranges of spectral types 
among the populations. Because the DM97 models produce
systematically older ages near the hydrogen burning limit 
compared to higher and lower masses (Luhman 1999), objects 
later than M4 in IC~348 and $\rho$~Oph are omitted (the spectroscopic samples 
in these regions are complete to M5). As illustrated in Figure~\ref{fig:ages1}, 
the $\rho$~Oph cloud core and the Trapezium interpreted with DM97 have 
similar median ages ($\sim0.4$~Myr).  The $5\arcmin\times5\arcmin$ core 
of IC~348 has an older median age (1-2~Myr) and lacks the
extremely young ages ($<0.5$~Myr) found in the other two clusters. 

For all three clusters, our data are consistent with very short durations for 
the star formation. The true range of ages may be more narrow still 
because most forms of observational error will artificially broaden 
the age distributions. For example, because the evolutionary
tracks are vertical on the H-R diagram at low masses for ages $\sim1$~Myr 
or greater, the ages of unresolved binary systems inferred from the
H-R diagram will be younger than would have been deduced for the individual 
components. Thus, a coeval population with a mixture of binary and single 
stars will appear to have a significant range of ages.

\subsubsection{The IMF}
\label{sec:imf}

\subsubsubsection{The Spectroscopic Sample}

The first step in constructing the Trapezium IMF was to
assemble a sample observed spectroscopically, within
the $140\arcsec\times140\arcsec$ NICMOS field towards the Trapezium,
and with $A_H\leq1.4$, which includes a majority of the Trapezium population
while remaining free of background stars (see \S~\ref{sec:member}).
Each binary system that is resolved in the {\it HST} data of Prosser et
al.\ (1994) but not in the NICMOS images is treated as a single star. 
The partially resolved pair TCC075 and TCC077 is also treated as one star.
The unresolved G-type companion to the B star 
119 is ignored. All remaining sources have separations greater
than $0\farcs7$, a physical scale that is comparable to those resolved 
in ground-based studies (1-$2\arcsec$) of IC~348 and $\rho$~Oph.
The resulting IMFs should be similar to the primary star mass functions,
with the exception of binary companions at large separations, which will 
be included in the IMF under the above prescription.

For the six OB stars, we adopt the masses from Hillenbrand (1997). 
In studies of the Orion Nebula Cluster, Hillenbrand (1997) 
and Hillenbrand \& Hartmann (1998) found evidence for mass 
segregation for stars above 5~$M_{\odot}$. They suggest that the 
concentration of the massive stars in the center of the cluster is likely
primordial rather than due to dynamical effects. We do not account for
this segregation in the IMFs shown in this paper. 
In such a correction, the four stars above 5~$M_{\odot}$ would be normalized 
to the number of stars in a reddening limited sample from the entire 
Orion cluster, which would reduce the counts in the most massive bins of 
the IMF by an order of magnitude.

The DM97 and B98 IMFs for the Trapezium spectroscopic sample are shown
as the dashed histogram in Figs.~\ref{fig:imf1} and \ref{fig:imf2}.

\subsubsubsection{Sources Without Spectra}
\label{sec:complete}

In a similar manner as LR98, LRLL, and LR99, we estimate
masses for the faint sources that lack spectra and are likely members
and add them to the IMF. As previously discussed, we expect all objects
within the reddening limit of the IMF calculation to be cluster members. 
This correction to the spectroscopic sample
will be well-defined in terms of mass and 
extinction because 1) our photometry encompasses the wavelengths where cool, 
low-mass objects are luminous, 2) the NICMOS color is insensitive to intrinsic
spectral types, 3) the IR photometry reaches reddened objects easily, 
4) the directions of increasing reddening and decreasing mass are nearly
perpendicular in an IR color-magnitude diagram (unlike in the optical), and
5) the abrupt increase in extinction between the Trapezium population and the 
background stars behind the molecular cloud assures us that there is no field 
star contamination in a reddening limited sample (see \S~\ref{sec:member}).
For this correction, we include all sources that lack spectral types, are
detected at both m$_{110}$ and m$_{160}$, and have $A_H\leq1.4$. We omit
the four sources with anomalously blue colors of m$_{110}-{\rm m}_{160}<0.5$.
The unresolved binaries from Prosser et al.\ (1994) are treated as one star.
For the sources with NICMOS photometry, reddenings are derived from 
m$_{110}-{\rm m}_{160}$ as described in \S~\ref{sec:colors}. In addition, 
there are a few sources that lack both spectral types and NICMOS photometry. 
Object 46 is not detected in the NICMOS images and hence is probably too red 
to be included in the IMF sample. Star 136 is saturated in the NICMOS data, 
but has $I$ and $K$ measurements that are similar to those of 113, 130, 187, 
and 238, which have inferred masses of 2-3~$M_{\odot}$. We therefore place 
136 in the mass bin centered at log~$M=0.45$. Sources 120, 127, 153, 154, 
and 191 are too close to the OB stars to be measured by NICMOS and 145 falls 
within a diffraction spike. Because only $K$-band data are available for 
these six objects, we will adopt a reddening of $A_K=0.3$, which is typical 
for the center of the Trapezium. 

By combining the dereddened photometry with the derived ages, 
we can estimate masses for the sources that lack spectra.
We describe such a derivation of masses first with the DM97 models.
The distribution of DM97 ages for Trapezium stars with
spectral types earlier than M1 
(omitting the OB stars) should be representative of the entire population 
(\S~\ref{sec:ages}).  We normalize this star formation history to the
total number of objects later than M0 $--$ stars classified later than M0 or
faint stars lacking spectral types. From this distribution,
we subtract the histogram of ages for
stars that fall in the spectroscopic sample that are later than M0;
the resulting distribution should reflect the ages of the objects without 
spectral types (see Figure~\ref{fig:ages2}).
For each such object, we estimate a mass by 
combining an age randomly drawn from the derived 
distribution with the dereddened
m$_{160}$ (or $K$) photometry, distance modulus, DM97 models,
and bolometric corrections.  After repeating this procedure 10 times
for each source and computing the average of the 10 resulting masses, we 
arrive at masses that are added to the DM97 spectroscopic
IMF, producing the final DM97 mass function in Figure~\ref{fig:imf1}.
At times, the modeling produced 
masses falling below the lower limit of the plot of the IMF (log~$M=-1.85$).
For instance, the faintest unreddened object in the
color-magnitude diagram in Figure~\ref{fig:hjh} is source 284 
(m$_{160}=17.1$, m$_{110}-{\rm m}_{160}=0.7$), which should have a mass of
only $\sim0.01$~$M_{\odot}$ at the median age of 0.4~Myr for the Trapezium.

Because the evolutionary models of B98 do not include ages younger than
1~Myr, we cannot derive a distribution of B98 ages for the entire Trapezium
Cluster, and thus we cannot estimate B98 masses for the sources lacking
spectra in exactly the same manner as done with the DM97 models. 
Instead, we adopt one age for all sources lacking spectral types
in estimating their B98 masses. The appropriate age is the median with
these models of $\sim1$~Myr, as indicated in Figure~\ref{fig:hr}.
This method should be equivalent to that used with DM97 for the following 
reasons. First, the mass bins that we have selected for the B98 IMF are so wide 
that the adopted age has little effect on the mass bin that an object falls in.
We have also used the DM97 models to test how the mass function is affected by
the adoption of a single age for all sources rather than a distribution of ages.
We derived DM97 masses for the sources lacking spectra first by randomly drawing
ages from the star formation history, as done in the previous paragraph, and 
then by adopting the median DM97 age of 0.4~Myr for all. We find that the
resulting mass functions are the same within the counting uncertainties. 
Thus, we expect that adopting the median of 1~Myr in estimating the
B98 masses should produce a mass function that is similar to the one we 
would find if we were able to apply a full distribution of ages. 
After accounting for these sources that lack spectra, we
arrive at the final form of the B98 IMF in Figure~\ref{fig:imf2}; this mass
function includes a population of $\sim50$ likely brown dwarfs. 

The above addition to the Trapezium IMF should be incomplete at the lowest
masses because older or reddened brown dwarfs can fall below the detection
limit of the NICMOS photometry. For instance, a brown dwarf at a mass of 
0.02~$M_{\odot}$ and an age of 3~Myr would appear at the detection limit
of $H=17$. To estimate the completeness of the substellar IMF, we construct
a population of brown dwarfs that is distributed uniformly across the 
mass bins from log~$M=-1.45$ to $-1.65$ and log~$M=-1.65$ to $-1.85$.
For each mass, an age is randomly drawn from the distribution of ages 
that is representative of the Trapezium, as performed earlier in this section.
An apparent $H$-band magnitude is inferred for each mass and age from the 
models of DM97. The combined distribution of magnitudes predicted for each mass
bin is then compared to the limits of the photometry. We find that the 
mass bins from log~$M=-1.45$ to $-1.65$ and log~$M=-1.65$ to $-1.85$
are $\sim90$\% and $\sim60$\% complete, respectively. We do not correct
for this incompleteness in Figs.~\ref{fig:imf1} and \ref{fig:imf2} and
instead wait for deeper photometry and spectroscopic confirmation of 
some of these brown dwarfs.  However, this simulation suggests that the 
mass function in the Trapezium could be flat down to 0.01-0.02~$M_{\odot}$, as
found for the somewhat older cluster $\sigma$~Ori (Zapatero Osorio et 
al.\ 2000).

\section{The Initial Mass Function in Different Environments}
\label{sec:compare}

\subsection{Star-Forming Clusters}

Uncertainties in the theoretical evolutionary tracks can
influence the shape of the IMF derived for young clusters.
However, in comparing clusters with similar star formation
histories, these effects will distort the derived IMFs in
similar ways. That is, issues with the theoretical tracks 
will cancel to first order. Therefore,
we use the DM97 tracks for comparisons among the three
clusters we have studied in depth, IC~348, $\rho$~Oph, and the Trapezium.
Although we believe the B98 tracks are more accurate, those of 
DM97 have the advantages of extending to younger ages, and of providing
the reader with a straightforward comparison with our previous
work, which used the DM97 tracks. 

\subsubsection{The IMFs in IC~348 and $\rho$~Oph}

LRLL used a combination of spectroscopy and photometry to construct
the IMF for members of the $5\arcmin\times5\arcmin$
core of IC~348. Luhman (1999) obtained
spectra of additional low-mass candidates for which only
photometry had been available previously, identifying five new members
and several background stars. To compute the 
IMF and examine the completeness, we consider sources
from these two references and with $A_V<7$, a sample that
includes all but two of the known cluster members towards the core.

A recent proper motion study of IC~348 (Scholz et al.\ 1999) has provided
membership probabilities for the foreground, background, and cluster 
populations within a square degree surrounding IC~348 at $R<18$.
These data are not definitive for distinguishing between cluster 
members and background stars. Many stars have high probabilities of belonging 
to both populations and several stars that are clearly cluster members by their 
photometry and spectra are given low cluster membership probabilities. 
Foreground stars are more confidently identified. 
Sources 77 and 121 from LRLL are probably in the foreground,
which is consistent with the low reddenings implied by their colors and the
lack of emission lines or IR excess. Both stars were considered cluster members 
by LRLL, and 77 was included in the IMF calculation of the cluster core. We 
omit 77 from the IMF shown here. The cluster memberships of the vast majority 
of the remaining sources in LRLL and Luhman (1999) are established by 
properties such as reddened colors and spectra, emission lines, IR excess 
emission, and gravity-sensitive spectral lines.

We took the masses of the cluster members observed
spectroscopically from LRLL and Luhman (1999). To determine
the IMF, we must add any likely cluster members that lack spectral types.
Three objects from LRLL (96, 230, 248) 
appear to have late-type IR spectra ($>$K5), and therefore cannot be in 
the background because of their brightness.  The reddening of these 
three stars is low, so they could be in the foreground. But because they 
were not identified as foreground stars in the proper motion study of Scholz et 
al.\ (1999) and because of the low probability of foreground star contamination 
towards the small area of the core (Herbig 1998), 
we take them to be cluster members for this analysis. 
IC~348 does not have a thick background
molecular cloud; thus, background stars can appear in the
photometry at low reddenings. However, member brown dwarfs
have higher values of $R-I/J-H$ and $I-K/J-H$ than
reddened field stars (see, e.g., Luhman 2000). By combining $R$
and $I$ photometry (Luhman 1999) with $JHK$ data for the cluster
core (LRLL), we identify and reject the background stars in the core down to 
very faint limits ($H\lesssim16.5$) and identify five additional
brown dwarf candidates listed in Table~2. Source 435 falls below the locus
of cluster members in a diagram of $R-I$ vs.\ $I$ (Luhman 1999), but this
is likely the result of blue excess emission, as suggested by the abundance
of strong emission lines in its optical spectrum.
Reddenings for these eight objects are measured from $J-H$ assuming an
intrinsic color of 0.7. 
Masses are estimated by combining the dereddened $H$ magnitudes with 
the DM97 models for an assumed age of 3~Myr.
The resulting IMF is given in Figure~\ref{fig:imf1}. 
 
The magnitude ($H\sim16.5$) and reddening
limits ($A_V<7$-8) are similar between the completeness corrections for the 
Trapezium and IC~348. Since the differences in distances 
and ages tend to cancel, we expect the substellar IMF in IC~348 to have a 
similar completeness level as described earlier for the Trapezium.

The mass function for $\rho$~Oph is taken directly from LR99 and may still be 
slightly incomplete in the substellar range. For this cluster, the results 
from our analysis have been shown to be in excellent agreement with analyses 
based on photometry alone (e.g., Williams et al.\ 1995; Comer\'on, Rieke, \&
Rieke 1996). This agreement is improved if these earlier analyses are 
corrected approximately to expectations for the DM97 tracks (which would tend 
to reduce the portion of very low mass objects slightly). This agreement 
supports our use of a combination of spectroscopic and photometric methods 
to arrive at complete IMFs, since there appear to be no significant systematic 
differences in the results of the two approaches. 

\subsubsection{Comparison of IMFs in IC~348, $\rho$~Oph, and the Trapezium}

It can be difficult to make meaningful comparisons of reported
mass functions because of differing methodology and the wide range of 
environments investigated (Scalo 1998). For instance,
the field star mass function of Reid \& Gizis (1997)
(updated by Reid et al.\ 1999) differs significantly
from the mass functions determined in other 
studies (Kroupa 1995a, 1995b; Gould et al.\ 1997).  
In Figure~\ref{fig:imfprev} the IMFs reported recently for the Pleiades 
(Bouvier et al.\ 1998) and the field (Reid et al.\ 1999) are similar, but
these mass functions do not match that of
Orion (Hillenbrand 1997). One would have expected the mass
functions in the Pleiades and Orion to be similar, since
they are both dense clusters and differ primarily in age.

At least some of these variations arise because different techniques (colors
vs.\ spectral types) and evolutionary models (DM97 vs.\ B98) are often
employed in converting data to masses.
To illustrate, we compare our IMF for the center of the Trapezium 
($D=140\arcsec$) to the results reported by Hillenbrand (1997) for the larger 
Orion Nebula Cluster ($D=18\arcmin$). The IMF of Hillenbrand is based on the 
DM94 tracks and has a small excess and a deficit of stars from 0.1-0.25 and 
0.4-1~$M_{\odot}$, respectively, relative to our DM97 IMF shown in 
Figure~\ref{fig:imf1}. This difference persists when the comparison is
restricted to stars in common between the two studies. However, when
we adjust the luminosities estimated by Hillenbrand to
the distance that we have adopted and derive masses
for her data from the models of DM97 rather than DM94, the
revised version of Hillenbrand's IMF agrees well with our measurements.
While theoretical tracks for low-mass stars are mostly vertical on the
H-R diagram for ages $\gtrsim1$~Myr, they do exhibit significant dependence on
both temperature and luminosity at younger stages,
hence the dependence of mass on the estimated luminosity for
the Trapezium. Unfortunately, the deficit of stars at 0.4-1~$M_{\odot}$
and the peak at 0.2~$M_{\odot}$ in the DM94 IMF of Hillenbrand have been
referred to as real features of the IMF (Sirianni et al.\ 1999).

Whereas the results of Hillenbrand (1997) used the models of DM94 and were
complete to 0.1~$M_{\odot}$, the Orion IMF of Hillenbrand \& Carpenter 
(2000) was based on the models of DM97 and reaches 0.02~$M_{\odot}$. After
comparing the IMF of Hillenbrand \& Carpenter (2000) (their Figure 16) to
our DM97 IMF (Figure~\ref{fig:imf1}), we find that they are in agreement; they
both show a peak near 0.2~$M_{\odot}$ followed by a slow decline and 
flattening into the brown dwarf regime. 

Our studies of IC~348, $\rho$~Oph, and the Trapezium cluster
use homogeneous observational approaches and interpret
the data with similar analyses using one
theoretical foundation. The similarity in ages among these
clusters causes most potential sources of systematic error
to cancel to first order. Therefore, we can reliably
search for variations in the low-mass
IMF across two orders of magnitude in stellar density ($\rho$~Oph, 
$n=0.2$-$1\times10^3$~pc$^{-3}$; IC~348, $n=1\times10^3$~pc$^{-3}$;
Trapezium, $n=1$-$5\times10^4$~pc$^{-3}$) and the
accompanying range in star formation efficiency.
As seen in Figure~\ref{fig:imf1}, the IMFs in 
IC~348, $\rho$~Oph, and the Trapezium are quite similar. With
the evolutionary models of DM97, the IMFs are
characterized by a slow increase
from substellar masses to $\sim0.25$~$M_\odot$, and a drop with a slope of 
$\sim0.7$ (Salpeter is 1.35) from 0.25 to 3~$M_\odot$. The mass
functions from brown dwarfs to the lowest mass stars are similar between
the Trapezium and $\rho$~Oph. This comparison has the highest
weight because of the very similar ages of the clusters.
To first order, the IMF for IC~348 is also similar to those for the
other two clusters. To second order, there appear to be proportionately fewer
brown dwarfs in the core of IC~348. This tendency is only of modest statistical
significance (compare the demonstration by Elmegreen 1999a of the effects of 
statistics on the high mass IMF).
A better comparison of the substellar mass functions requires spectroscopy of
the brown dwarf candidates that comprise the completeness corrections in these 
clusters to confirm their cluster membership and measure more precise masses. 
In addition, the spectroscopy survey begun by Luhman (1999) of all of IC~348
should eventually provide much better number statistics ($\times3$) for
comparison to $\rho$~Oph and the Trapezium.

Above a few solar masses, the IMFs in the clusters IC~348, $\rho$~Oph, and 
the Trapezium are the same within the uncertainties. Better number statistics 
can be achieved in richer, distant populations.  For instance, in observations 
of clusters in the Large Magellanic Cloud, Hunter et al.\ (1997) and 
Massey \& Hunter (1998) found that the IMF above 1~$M_{\odot}$ was invariant 
with cluster density over two orders of magnitude. Scalo (1998) 
found that the various IMFs reported in the literature for this mass range
cannot be easily reconciled, even among studies of the same regions. 
If the variations in the mass function are real, they do not appear to
depend on metallicity, stellar density, or Galactocentric radius (Scalo 1998).
Elmegreen (1999a, 1999b) suggests that much of the variation in the measured
slopes of the IMF arises from statistical fluctuations 
and that there may be a common IMF at intermediate masses.

\subsection{Star-Forming Clusters, Young Open Clusters, \& the Field}
\label{sec:true}

While no set of theoretical evolutionary models agrees with observational 
tests at all ages, masses, and metallicities, White et al.\ (1999) and
Luhman (1999) concluded that the models of B98 are the most consistent with the 
observational constraints available for very young low-mass stars. In addition,
Luhman (1999) found that a temperature scale intermediate between those of M 
dwarfs and giants produced the best agreement between the B98 calculations and 
observations.  By combining this temperature scale with the models of B98, 
we should arrive at the most accurate IMF currently possible for a young
stellar population.

The DM97 and B98 IMFs for the Trapezium and mass functions of the Pleiades
and the field are compared in Figure~\ref{fig:imf2}. With the DM97 models, 
the IMF in the Trapezium (and the other star-forming clusters) 
peaks near $\sim0.25$~$M_{\odot}$. In the more accurate B98 IMF, the peak 
shifts to higher masses and the shape is quite similar to the results 
in the Pleiades and the field; the differences between the previous Orion IMF
and the other two regions shown in Figure~\ref{fig:imfprev} are now removed. 
In addition, Barrado y Navasc\'ues et al.\ (2000) have recently reported that
the stellar IMF for the M35 open cluster is similar to that of the Pleiades. 
We find that the data for the Trapezium, the Pleiades and M35 open clusters,
and the field are all consistent with the same mass function, one that is flat
or slowly rises from substellar masses, and rolls over between 0.6 and 
1.0~$M_\odot$ into a falling power law toward high masses. 
The slope of the Trapezium B98 IMF from log~$M=-0.25$ to $-1.45$
(0.56-0.035~$M_{\odot}$) is $\sim-0.3$ (Salpeter is 1.35). 
As illustrated in Figure~\ref{fig:imf2}, because of the more pronounced 
peak in the Trapezium IMF compared to the Pleiades (which could be a second 
order systematic error from the methods and models), the computed slope for
the Trapezium low-mass IMF is sensitive to the mass limits that are selected.

It is not surprising 
that the IMFs of the Trapezium and the Pleiades are similar since the Pleiades 
is a rich open cluster that was probably much like the Trapezium when it was
younger. Furthermore, the similarity of these cluster IMFs to the mass function 
in the field is consistent with the suggestion that the field is predominantly 
populated by stars that formed in clusters (Lada, Strom, \& Myers 1993). Since 
we might have expected a high degree of similarity among the mass functions, 
the observed agreement can alternately be taken as a demonstration that the 
methods for deriving the mass functions are consistent. 

The data for the star-forming clusters and the young open clusters are most 
useful for studying the IMF when combined to complement each other. Because 
the stars in clusters like the Pleiades and M35 are free of excess emission and 
significant reddening, temperatures
and luminosities are more accurately measured as compared to very young stars.
In addition, because the open cluster members are near the main sequence, 
the theoretical models should be more robust and provide better mass estimates. 
These open clusters are also very rich and provide excellent number statistics,
particularly in M35. Thus, the
detailed structure of the IMF is more readily determined in regions like the 
Pleiades and M35. On the other hand, as noted by Bouvier et al.\ (1998), the  
IMF measured for the Pleiades may be a lower limit at low masses because
of observational incompleteness and possible mass segregation. 
The substellar population is more luminous in 
clusters at the age of Orion and significant dynamical evolution has not 
occurred, allowing the measurement of the IMF down to much lower
masses than possible in the open clusters. From the work presented 
here, we find no obvious decline in the density of sources down to the 
detection limit of 0.02~$M_{\odot}$ and probably, with a correction for 
incompleteness (\S~\ref{sec:complete}), down to 0.01-0.02~$M_{\odot}$.

\subsection{Comparison with Other Studies}

The three young clusters in our study are by a significant margin the 
most thoroughly studied. However, the information available on other 
similar regions appears to be consistent with our results.
Comer\'on et al.\ (1996) used photometric techniques to
find a similar flat low-mass IMF to below the stellar limit in NGC 2024.
Comer\'on, Rieke, \& Neuh\"auser (1999) showed
that a flat IMF is also consistent with their data on Cham~I. 
There are indications in a few older open 
clusters for a deficiency of low-mass members (e.g., Hawley, Tourtellot, \& 
Reid 1999). However, for all of these clusters, there are concerns about the 
possible roles of dynamical evolution and mass segregation, about 
distinguishing members from background stars, and about the completeness of 
the known membership lists given the fading of the low-mass objects to or 
below the detection limits achieved in the near-IR and X-ray regions used 
for their identification. The available data for young open clusters are, in 
our view, all consistent with a common IMF of the form we have derived for 
the IC~348, $\rho$~Oph, and the Trapezium. 

We can also compare with recent estimates of the IMF in 
globular clusters and the bulge. Paresce \& De Marchi (1999) concluded that 
the data on various globular clusters are consistent with a log-normal IMF 
with a characteristic mass of 0.33~$M_\odot$. Towards the Galactic bulge, 
Zoccali et al.\ (1999) report an IMF with a slope of 0.33 from
0.15-0.5~$M_{\odot}$ and one similar to that of Salpeter ($\alpha=1.35$) from 
0.5-1~$M_{\odot}$. These estimates, particularly the one for globular clusters,
appear to differ significantly from the form of the IMF we find for young 
clusters and the field. This difference suggests that there may be a detectable 
variation in the IMF if the star forming conditions are changed sufficiently. 
Although the conditions that prevailed for globular clusters are now
impossible to determine observationally, this result should encourage searches 
for variations in other environments. 

\subsection{Implications for Theories of the IMF}

Given the shape of the low-mass IMF that we have measured, its approximate 
invariance at stellar masses among local regions of clustered star formation, 
and the constraints on the minimum mass of free-floating objects, what are 
the implications for theories of the origin of the mass function?

Previous studies indicate that the IMF flattens below 1~$M_{\odot}$ 
(Scalo 1999). Our observations of IC~348, $\rho$~Oph, and the Trapezium, 
with data for the Pleiades and M35 open clusters and the field, 
show that the effect occurs between 0.6 and 1~$M_\odot$, which
can be interpreted in terms of a characteristic mass of the IMF. 
When the origin of the IMF is attributed to random sampling of the 
hierarchical structure of molecular clouds (Henrikson 1986, 1991;
Larson 1992; Elmegreen 1997, 1999a), the IMF should flatten or turn over near  
the Jeans mass. This parameter is not 
likely to vary significantly among nearby star forming regions 
(Elmegreen 1999a), although such a conclusion could break down when the 
details involved in determining the Jeans mass are included (Myers 1998). Other 
theories (e.g., Silk 1995; Lejeune \& Bastien 1986; Murray \& Lin 1996) can
produce a flattening of the IMF without any consideration of the Jeans 
criterion (Scalo 1999); for example, a turnover has been suggested
to arise at or above the deuterium-burning mass (Shu, Adams, \& Lizano 1987).

The Jeans mass is a function of the gas
temperature and cloud-core pressure, and these properties in turn depend
on Galactocentric distance. It is unclear if the characteristic mass
we find in the IMF is indeed related to the Jeans mass without 
observing the low-mass IMF in clusters that span a wide range of Galactocentric
distances. The lower characteristic mass in globular clusters is encouraging 
that such variations will be found if a sufficiently large range of conditions 
can be probed.

Scalo (1998) has already stressed that there is little
evidence for a log-normal IMF. 
Data for globular clusters have been fit by a log-normal IMF
(Paresce \& De Marchi 1999), but because there are no 
constraints on the substellar mass functions in these regions, a true 
log-normal distribution cannot be verified.  The IMFs in the Trapezium,
the other young clusters, and the field are clearly not log-normal,
contrary to the predictions of some models (Klessen, Burkert, \& Bate 1998). 
Adams \& Fatuzzo (1996) also derived a log-normal IMF in the case that
a large number of independent variables determine the masses of stars. For
a smaller number of such variables, a power-law tail appeared at high masses,
which could match the form of the observed high-mass IMF. However, the IMF
in the Trapezium and other clusters is approximately flat down to 
0.01-0.02~$M_{\odot}$ and it is unclear whether the models of Adams \& 
Fatuzzo (1996) can reproduce such a dramatic deviation from log-normal form. 

The turnover in the IMF is fairly similar among clusters
which include a large range of stellar
densities and star formation efficiencies.
This result is consistent with theories where the
masses of stars are determined by processes
of accretion and outflows (Adams \& Fatuzzo 
1996), instabilities related to stellar winds (Silk 1995), or
hierarchical fragmentation (Elmegreen 1999a), but
difficult to reconcile with suggestions that the stellar masses
are controlled by dynamical interactions among stars
and protostars (Price \& Podsiadlowski 1995; Bonnell et al.\ 1997) 
or collision and coalescence of clumps (Lejeune \& Bastien 1986;
Murray \& Lin 1996). In these latter cases, the properties of the IMF,
such as the ratio of high to low-mass stars and the
turnover mass, should depend on stellar density.

The shape of the substellar mass function and the minimum mass observed for
free-floating objects are powerful constraints for theories of the IMF.
We find that brown dwarfs can form in moderately large 
numbers, whereas a minimum mass of $\sim0.1$~$M_{\odot}$ is predicted by some 
models of hierarchical fragmentation (Larson 1992).
Furthermore, it appears that free-floating objects can form at masses near
($\sim0.015$~$M_{\odot}$; Zapatero Osorio et al.\ 1999) and below
($\sim0.01$~$M_{\odot}$; this work and Zapatero Osorio et al.\ 2000) 
the deuterium burning limit (0.013-0.015~$M_{\odot}$; Burrows et al.\ 1997).
This behavior is contrary to the expectations of wind-limited 
models (Shu, Adams, \& Lizano 1987). In fact, most theories of star formation 
have difficulty in explaining the abundance and minimum mass of brown dwarfs 
that we observe (see Elmegreen 1999b). 

From mm-wave continuum observations in Serpens
(Testi \& Sargent 1998) and $\rho$~Oph (Motte, Andr\'{e}, \& Neri 1998), it
appears that the mass function of apparently pre-stellar clumps
is reminiscent of the stellar mass function in the field.
Umemoto et al.\ (1999) have identified clumps in $\rho$~Oph
using H$^{13}$CO$^+$ emission; the resulting clump
mass spectrum is similar to that of Motte et al.\ (1998)
but shifted toward higher masses by a factor of three to four.
The discrepancy may arise through modification of
mm-wave emission properties of grains in dense clumps
(Kruegel \& Siebenmorgan 1995). 

LR99 showed that the stellar IMF in $\rho$~Oph
determined with the DM97 tracks bears a
close resemblance to the clump mass function in the same
region determined by Motte et al.\ (1998).
Use of the more accurate B98 tracks shifts the
turnover mass in the stellar IMF upward. As shown in
Figure~\ref{fig:clumps}, the turnover lies above that
of the Motte et al.\ clump spectrum; it lies below
the turnover in the Umemoto et al.\ spectrum.
In both cases, the agreement is within a factor of
two. Given the potential errors in measuring
the clump masses, it is possible that we are
seeing evidence for a connection between the process
of cloud fragmentation and the IMF.
Better number statistics and mass accuracies and completeness to lower masses
are required to confirm this possible relation. In particular, the clump
mass function should be measured where the stellar IMF is
most distinctive, below the turnover mass and into the brown dwarf regime.

\section{Conclusion}

The shape of the low-mass IMF ($<1$~$M_{\odot}$) has remained uncertain 
because of several issues, the most important being incompleteness 
for faint low-mass stars and brown dwarfs. In observations summarized in this 
work, we have made reliable measurements of the IMF down to well below the
hydrogen burning limit and have presented a robust comparison of the low-mass 
IMFs in local regions of clustered star formation. Our results are as follows:

\begin{enumerate}

\item
Despite a range of two orders of magnitude in the density of
star formation, the {\it stellar} IMFs in the core of IC~348 
and the cloud core of $\rho$~Oph are the same as in the
Trapezium within the uncertainties. The IMFs in
the {\it brown dwarf} regime are also roughly similar, 
although more observations are required for a definitive comparison of the
substellar mass functions. 

\item
Data for the Trapezium, the Pleiades and M35 open clusters (Bouvier et 
al.\ 1998; Barrado y Navasc\'ues et al. 2000), and the field 
(Reid et al.\ 1999) are consistent with the same IMF; this mass function
is flat or slowly rising from the brown dwarf regime to 0.6-1.0~$M_\odot$, 
where it rolls over to a power law with a slope of $\sim1.7$, similar to or 
slightly steeper than the Salpeter value of 1.35.  The similarity 
of these mass functions is consistent with the suggestion that members of the 
field have formed predominantly in clusters.

\item
Whereas the IMF that characterizes young clusters and the field 
rolls over near 0.8~$M_\odot$ and is not log-normal, 
recent studies have found that the IMFs of 
globular clusters peak near 0.3~$M_\odot$ and can be described by a 
log-normal function (Paresce \& De Marchi 1999). As a result, there appear
to be fundamental differences in the IMF between
globular clusters and Galactic disk clusters.

\item
Using the high spatial resolution and sensitivity of NICMOS images,
we have penetrated the bright nebulosity of the Trapezium and identified a 
population of $\sim50$ likely brown dwarfs,
where the least massive candidate is $\sim0.01$~$M_{\odot}$ if the median
age of the Trapezium is assumed. 
Most theories of the IMF do not predict the formation 
of free-floating objects in significant numbers at such low masses. 
For instance, this low-mass population rules out a log-normal IMF in
this cluster, contrary to some theories of star formation.

\end{enumerate}

\acknowledgements
M. McCaughrean kindly provided unpublished $K$-band photometry.
We are grateful to F. Allard, I. Baraffe, and F. D'Antona
for access to their most recent calculations and thank E. Bergin, F. D'Antona, 
B. Elmegreen, and P. Myers for helpful discussions. 
Comments on the manuscript by J. Stauffer and B. Wilking are appreciated. 
K. L. was funded by a postdoctoral fellowship at the Harvard-Smithsonian 
Center for Astrophysics.  Portions of this work were supported by NASA 
grant NAGW-4083 under the Origins of Solar Systems program.

\appendix

\section{Method of Classification}

$K$-band spectral classification of young stars has been performed 
for L1495E, IC~348, and $\rho$~Oph, as described in detail 
by LR98, LRLL, and LR99. However, the classification of the Trapezium sample 
is more difficult than in the previous studies because of the combination of 
low spectral resolution ($R=800$) and bright nebular emission from the Orion
Nebula.  The $K$-band absorption lines that appear at this resolution are 
Br$\gamma$ (2.166~\micron), Na (2.206, 2.209~\micron), Ca (2.161, 2.163, 
2.166~\micron), and CO (2.294, 2.323, 2.353, 2.383~\micron), while weaker
lines of Mg (2.281~\micron) and Mg and Al (2.11~\micron) can sometimes be 
detected (LRLL). At a slightly higher resolution ($R=1200$), Mg and Al and 
several other metal lines can be measured accurately, providing better 
constraints on the spectral type and continuum veiling (LR99), 
where the veiling at 2.2~\micron\ is defined as 
$r_{K}=I_{2.2}({\rm IR~excess})/I_{2.2}({\rm star})$. The brightest 
transitions of H~I and H$_2$ in the Orion Nebula (Luhman, Engelbracht, \& 
Luhman 1998) fall near the photospheric absorption lines of Br$\gamma$, 
Mg/Al at 2.11~\micron, and several weaker metal lines between Na and Ca. 
Because the nebular emission varies on small scales, 
we could not accurately subtract the
emission lines from the stellar spectrum.
Thus, we have useful measurements of only Na, Ca, and CO.  
Because of the anti-correlation of Br$\gamma$ with the metal lines, 
it is an important line in the classification of G and early K stars,
and its loss makes more difficult distinguishing these stars
from heavily veiled M stars.

To classify the Trapezium stars, we compared each $K$-band spectrum to 
the others in the sample and organized most of the data into groups 
of spectra that appeared the same within the noise. 
The spectra in each group were then combined into a spectrum representative 
of that group. The composite spectra from the 14 groups are given in 
Figure~\ref{fig:spec} along with data for the BN object and source~n.
Wavelengths near H$_2$ 1-0~S(1) (2.122~\micron) and Br$\gamma$
(2.166~\micron) are not plotted because of contamination from nebular 
emission, except for BN, which is much brighter than the surrounding emission.
The composite spectra were classified by comparison to dwarf standards
from LR98 with various amounts of artificial veiling, and the resulting
classification was assigned to each star included in that composite. 
After deriving a spectral type for each star, we calculated the
percentage increase in the strength of the CO absorption over that
of the standard dwarf of that type.
The IR spectral types, veilings, and CO absorption strengths
are listed in Figure~\ref{fig:spec} and Table~1.  Spectra with very low 
signal-to-noise were not included in the composite spectra and have no IR
classification. A few stars exhibited steam absorption indicative of 
mid-to-late M types and were classified individually rather than combined 
into composites.

We adopted previously measured optical spectral types when available.
Otherwise, we used the IR classifications. For stars that have both IR and
optical spectra, the two spectral types generally agree with
within the uncertainties. This comparison is an important check 
of the IR classifications since the Trapezium stars are extremely young and 
show significant departures from dwarfs in their $K$-band spectra, as we
illustrate in \S~\ref{sec:gravities}. The optical classifications can also 
differ substantially among themselves; in these cases we adopted the optical 
type that was most consistent with the IR data. For stars with uncertain IR 
spectral types and no optical data, we adopted the average type for the 
optically classified objects with the same IR spectrum. An example is source 
207, which is classified as $\geq$G6 in the IR. Objects 257 and 266 have very 
similar IR spectra to 207 and have optical types of K0-K6 and G8-K2, hence we
adopted K0-K6 for 207. For object 114, the IR type is K4-M2. However, since the 
other five stars with the same IR spectra have spectral types covering a 
smaller range, K7-M2, we take this as the classification for 114. Object 119 
is a binary system where the components have optical spectral types of B5-B8 
and G0-G5.  At the spectral resolution of our $K$-band data, the B star should 
be featureless in the $K$-band except for Br$\gamma$ absorption. Hence, our 
spectrum of this unresolved system indicates a spectral type of G or K with 
continuum veiling, which arises from the B star rather than an IR excess in 
this case. 
For object 133, we find that the $K$-band spectrum implies 
a spectral type of K0-K7, whereas Hillenbrand (1997) reports an optical type 
of F2-F7. Spectra from two different nights confirm that we observed the 
correct star. We adopt the IR classification for this object, although it 
could be binary system where an F primary and a K secondary dominate in the 
optical and IR, respectively, similar to source 133.

\section{Colors and Extinctions}
\label{sec:colors}

To measure reddenings for the sources in Table~1, we examine
the various optical and IR colors that are available. We then determine
the reddening relation and intrinsic stellar colors for the NICMOS bands.
Standard dwarf colors are taken from the compilation
of Kenyon \& Hartmann (1995) for types earlier than M0 and from the young 
disk populations described by Leggett (1992) for types of M0 and later. 

The color excesses $E(V-I_C)$, $E(I_C-K)$, and $E(I_C-{\rm m}_{160})$ have been
computed for stars with spectral types by assuming the intrinsic colors are
those of standard dwarfs. The correlations among these excesses and 
m$_{110}-{\rm m}_{160}$ are then examined. Several
anomalous sources are very red in $V-I_C$ relative to 
m$_{110}-{\rm m}_{160}$ and $I_C-{\rm m}_{160}$, which is likely a result of 
cool companions. In addition, $V-I_C$ is prone to contamination from star spots 
and the derived extinction is sensitive to errors in spectral type for M stars
(Gullbring et al.\ 1998). The $E(I_C-K)$ excess and other colors associated 
with $K$ are systematically redder for a portion of the sample, suggesting 
the presence of IR excess emission at $K$. 
We find that the best reddening determination uses the $I_C-{\rm m}_{160}$ 
color, providing a long wavelength baseline for measuring
reddening while remaining relatively free of short or long wavelength excess 
emission. One disadvantage of this color is that the two bands were measured at 
different epochs and variability will increase the uncertainty in the color.
We note that the colors $I_C-J$, $I_C-H$, and $I_C-K$ 
may differ from the dwarf colors for young cool sources (Luhman 1999), 
in which case $I_C-{\rm m}_{160}$ would not be the ideal color for estimating 
reddening. However, this is not a problem here since there are few late M 
objects in our spectroscopic sample. 
For effective wavelengths of 8100~\AA\ (for M stars), 
1.60~\micron, and 1.65~\micron\ for $I_C$, m$_{160}$, and $H$, respectively, 
the extinction law of Rieke \& Lebofsky (1985) gives reddening relations of 
$A_I=3.32 A_H$, $A_{160}=1.06 A_H$, and $A_H=E(I_C-{\rm m}_{160})/2.26$.
For sources lacking $I_C$, reddenings 
are measured from m$_{110}-{\rm m}_{160}$
with the extinction relations and intrinsic colors derived below.
Saturated stars in the NICMOS images are dereddened to the standard
dwarf values of $V-I_C$ with $A_H=E(V-I_C)/2.23$. The reddenings in 
Table~1 are computed for the average spectral types of the adopted ranges.

As extinction increases, the effective wavelengths of the 
NICMOS band passes, particularly the very wide F160W, shift to longer 
wavelengths. We simulated the change in m$_{110}-{\rm m}_{160}$ as a function 
of reddening from $A_H=0$ to 4 assuming uniform transmissions for the
F110W (0.8-1.4~\micron) and F160W (1.4-1.8~\micron) filters and adopted the 
functional form of the reddening law found in Cardelli, Clayton, \& Mathis 
(1989).  The results of such modeling are independent of $R_V$ for bands 
longward of $V$. The simulated reddening relation
does depend on the shape of the intrinsic stellar spectrum, although the effect
is only noticeable at large extinctions. For instance, a reddening of 
$A_H=1.4$ leads to $E({\rm m}_{110}-{\rm m}_{160})=1.20$, 1.30, and 1.41 
for effective temperatures of 3000, 4000, and 6000~K, where a synthetic 
spectrum of Allard, Alexander, \& Hauschildt (1998) was used for 3000~K and 
blackbodies were assumed for 4000 and 6000~K.  As shown in the color-magnitude 
diagram in Figure~\ref{fig:hjh}, most sources in the NICMOS data
have reddenings of $A_H\lesssim1.4$, thus the reddening relation for 
$T_{\rm eff}=4000$~K should be sufficient for all sources.  In 
Figure~\ref{fig:jhhk}, this reddening vector is plotted with m$_{160}-K$ versus 
m$_{110}-{\rm m}_{160}$ using the $K$-band data of MS for the central square 
arcminute of the Trapezium. The simulated colors redden more slowly with 
additional extinction in a manner consistent with the reddening vector implied 
by the embedded stars in Figure~\ref{fig:jhhk}.

The origin of a reddening vector in Figure~\ref{fig:jhhk} will correspond 
to the intrinsic colors of a given star, or star-disk system in the case of a 
classical T~Tauri star. One example of such a vector is shown. 
After dereddening m$_{110}-{\rm m}_{160}$ with 
extinctions derived from $E(I_C-{\rm m}_{160})$, we find
intrinsic NICMOS colors of $0.8\pm0.3$ independent of K and M spectral types.
The large scatter in this dereddened color is not surprising considering the
observational errors in m$_{110}-{\rm m}_{160}$ and the uncertainties
in the $E(I_C-{\rm m}_{160})$ reddening from possible variability between 
measurements of $I_C$ and m$_{160}$. The intrinsic NICMOS colors of these young
stars can also be estimated by examining the locus of sources in 
Figure~\ref{fig:hjh}. A large number of stars have colors of 
m$_{110}-{\rm m}_{160}=1$-1.4. If these are the least reddened objects in the 
cluster and if the Trapezium has a minimum extinction of $A_V\sim2.4$ 
(Herbig \& Terndrup 1986), then an intrinsic color of $\sim0.8$ is again 
implied. The blue boundary of the locus in Figure~\ref{fig:hjh} is fairly 
constant for all magnitudes shown, supporting the notion that 
m$_{110}-{\rm m}_{160}$ does not depend significantly on spectral type. 
Assuming an intrinsic color of 0.8, the simulated reddening relation for a 
star of $T_{\rm eff}=4000$~K is 
$A_H=-0.546+0.550({\rm m}_{110}-{\rm m}_{160})+0.179({\rm m}_{110}-{\rm m}_{160})^2$.

Several close pairs in the optical {\it HST} images of Prosser et al.\ (1994) 
are unresolved in the other data. For these systems, we combined the 
optical photometry of the two components and treated them as one object. When
$V-I_C$ was not available for one component, $V-I_C$ for the other
star was adopted for the system. The optical data for ID488a assigned to 123 in
Table~1 probably applies to both 116 and 123. Since we have no $V$ or 
NICMOS data for these two stars, the reddening is measured by assuming that each
star has the $I_C-K$ color of the composite system.
The optical photometry for ID524 also applies to both 165 and 169. 
Extinctions were computed from the NICMOS colors for these two stars.
The extinction for object 49 was estimated with $I_C-K$, the only
color available.

\newpage

\begin{figure}
\caption{
NICMOS F110W image of the Trapezium Cluster ($140\arcsec\times140\arcsec$).
The display range is from 0 to 0.3 mJy per pixel. 
East is left and north is up. 
}
\label{fig:f110}
\end{figure}
\clearpage
 
\begin{figure}
\caption{
NICMOS F160W image of the Trapezium Cluster ($140\arcsec\times140\arcsec$).
The display range is from 0 to 0.3 mJy per pixel. 
East is left and north is up. 
}
\label{fig:f160}
\end{figure}
\clearpage
 
\begin{figure}
\epsscale{0.85}
\plotone{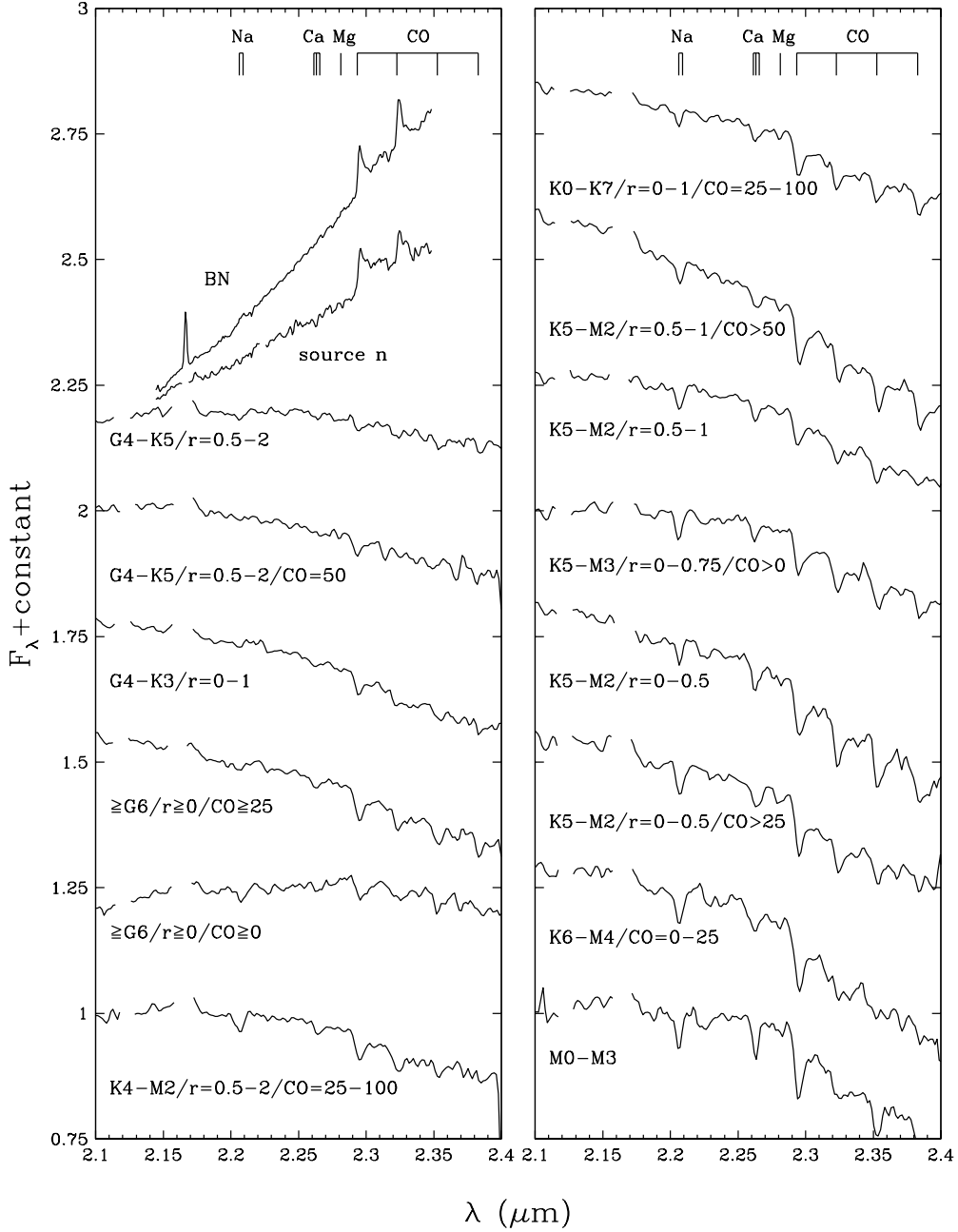}
\caption{
Composite $K$-band spectra of sources in the Trapezium Cluster.  The 
spectra of BN and source~n have $R=1200$ and are normalized at 2.29~\micron,
while the remaining data have $R=800$ and are normalized at 2.2~\micron.
}
\label{fig:spec}
\end{figure}
\clearpage
 
\begin{figure}
\epsscale{0.85}
\plotone{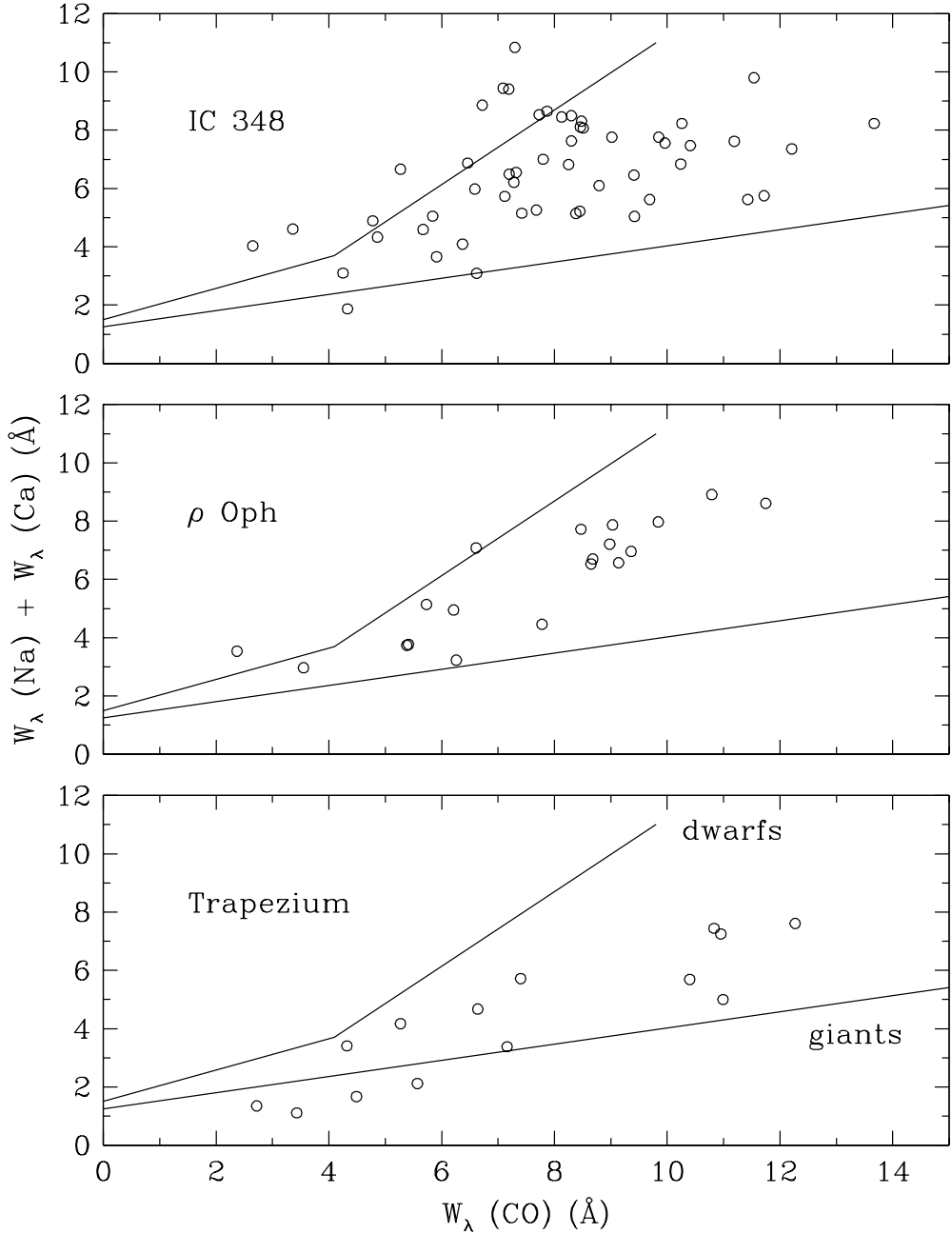}
\caption{
Equivalent widths of CO(2,0) versus the sums of the widths of Na and Ca 
in $K$-band spectra of young stars in the core of 
IC~348 ($5\arcmin\times5\arcmin$; LRLL), the cloud core of $\rho$~Oph
(LR99), and the Trapezium Cluster. The equivalent widths for the
Trapezium stars are measured from the composite spectra in 
Figure~\ref{fig:spec}. The solid lines represent standard dwarfs ($<$M5V) 
and giants ($<$M0III) (LR98). Significant $K$-band continuum 
veiling can occur in very young stars, which dilutes the equivalent widths
and moves the stars towards the origin. Typical measurement errors in the
equivalent widths are 0.3-1~\AA.
}
\label{fig:width}
\end{figure}
\clearpage
 
\begin{figure}
\plotone{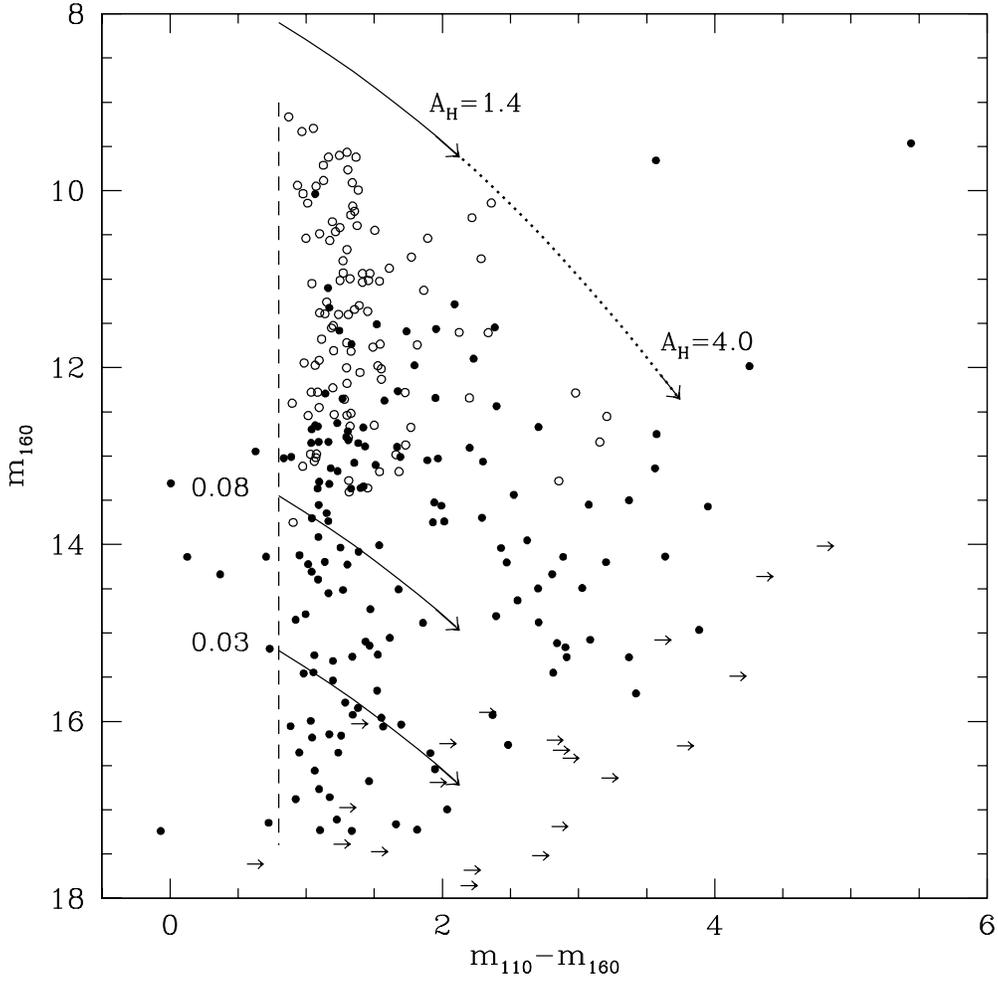}
\caption{
m$_{110}-{\rm m}_{160}$ vs.\ m$_{160}$ for the Trapezium Cluster
($140\arcsec\times140\arcsec$). Stars with spectral types are represented
by the open circles. The average intrinsic color
({\it dashed line}) is largely independent of spectral type for K and M stars. 
The upper curve represents the reddening vector in the NICMOS bands. 
Reddening vectors from $A_H=0$-1.4 are also shown for 0.08 and 
0.03~$M_{\odot}$ at an age of 0.4~Myr (D'Antona \& Mazzitelli 1997). The IMF 
is constructed from sources with $A_H\leq1.4$. 
}
\label{fig:hjh}
\end{figure}
\clearpage
 
\begin{figure}
\epsscale{0.85}
\plotone{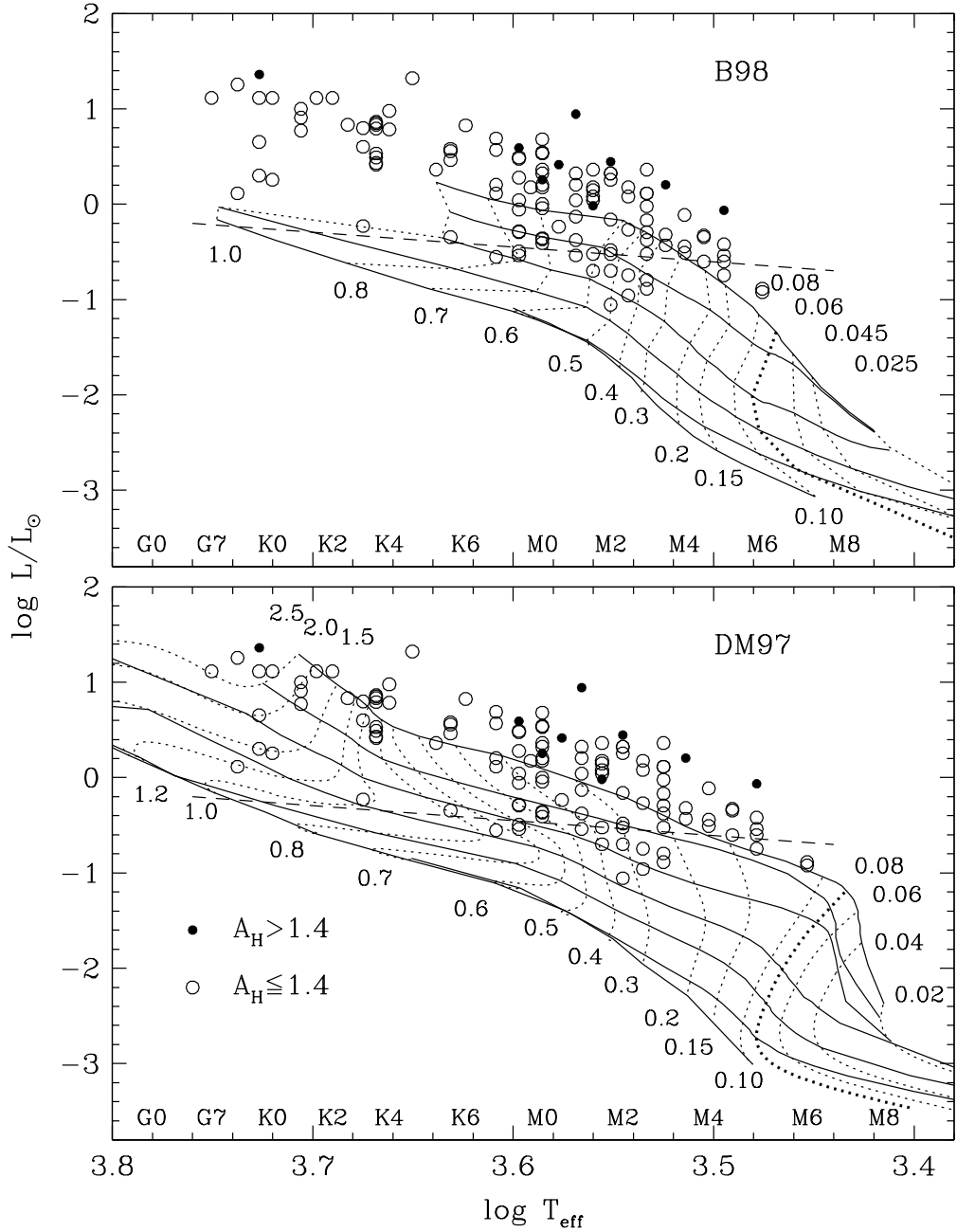}
\caption{
H-R diagram for the Trapezium Cluster ($140\arcsec\times140\arcsec$).
The theoretical evolutionary models of B98 (upper) and DM97 (lower) are 
shown, where the horizontal solid lines are isochrones representing ages 
of 0.3 (not available for B98), 1, 3, 10, 30, and 100~Myr and the main 
sequence, from top to bottom.
The dashed line in the H-R diagram represents a dereddened magnitude of 
$K=12$, above
which the spectroscopic sample is representative for $A_H\leq1.4$
(see Figure~\ref{fig:hjh}).  The M spectral types have been converted to 
effective temperatures with temperature scales that are compatible with 
each set of evolutionary models (Luhman 1999); a dwarf scale for DM97 and 
a scale intermediate between dwarf and giants for B98.
}
\label{fig:hr}
\end{figure}
\clearpage
 
\begin{figure}
\epsscale{0.85}
\plotone{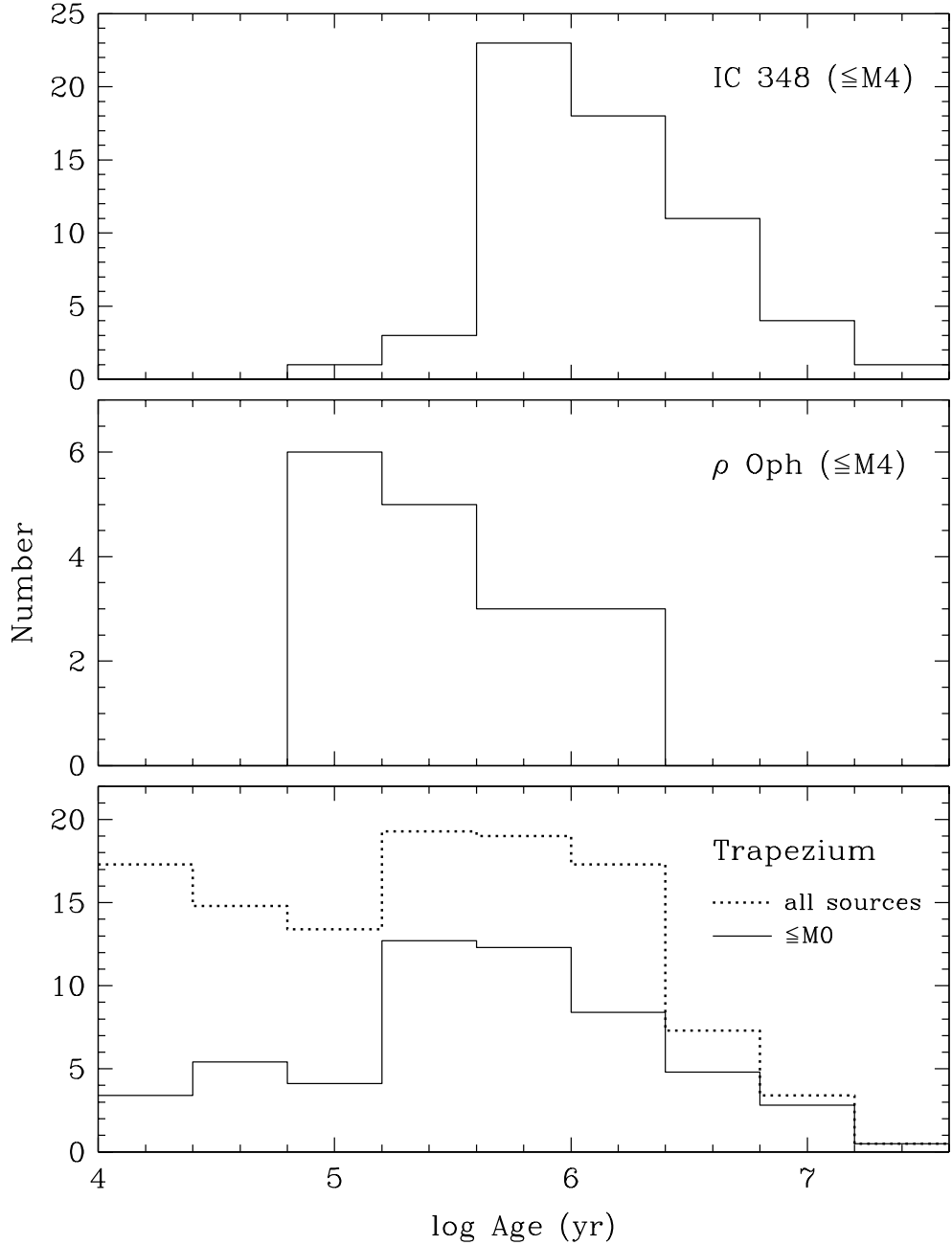}
\caption{
Distributions of ages inferred from the evolutionary models of DM97 
for the core of IC~348 ($5\arcmin\times5\arcmin$; LRLL), the cloud core of 
$\rho$~Oph (LR99), and the Trapezium Cluster ($140\arcsec\times140\arcsec$).
Objects with spectral types later than M4 in IC~348 and $\rho$~Oph have been 
omitted because the models of DM97 imply older ages for sources near the
hydrogen burning limit. For the Trapezium, the distributions of ages for
the entire spectroscopic sample ({\it dotted line}) and for stars M0 and earlier
({\it solid line}) are shown, where the latter should be representative of the
stellar population. 
}
\label{fig:ages1}
\end{figure}
\clearpage

\begin{figure}
\epsscale{0.8}
\plotone{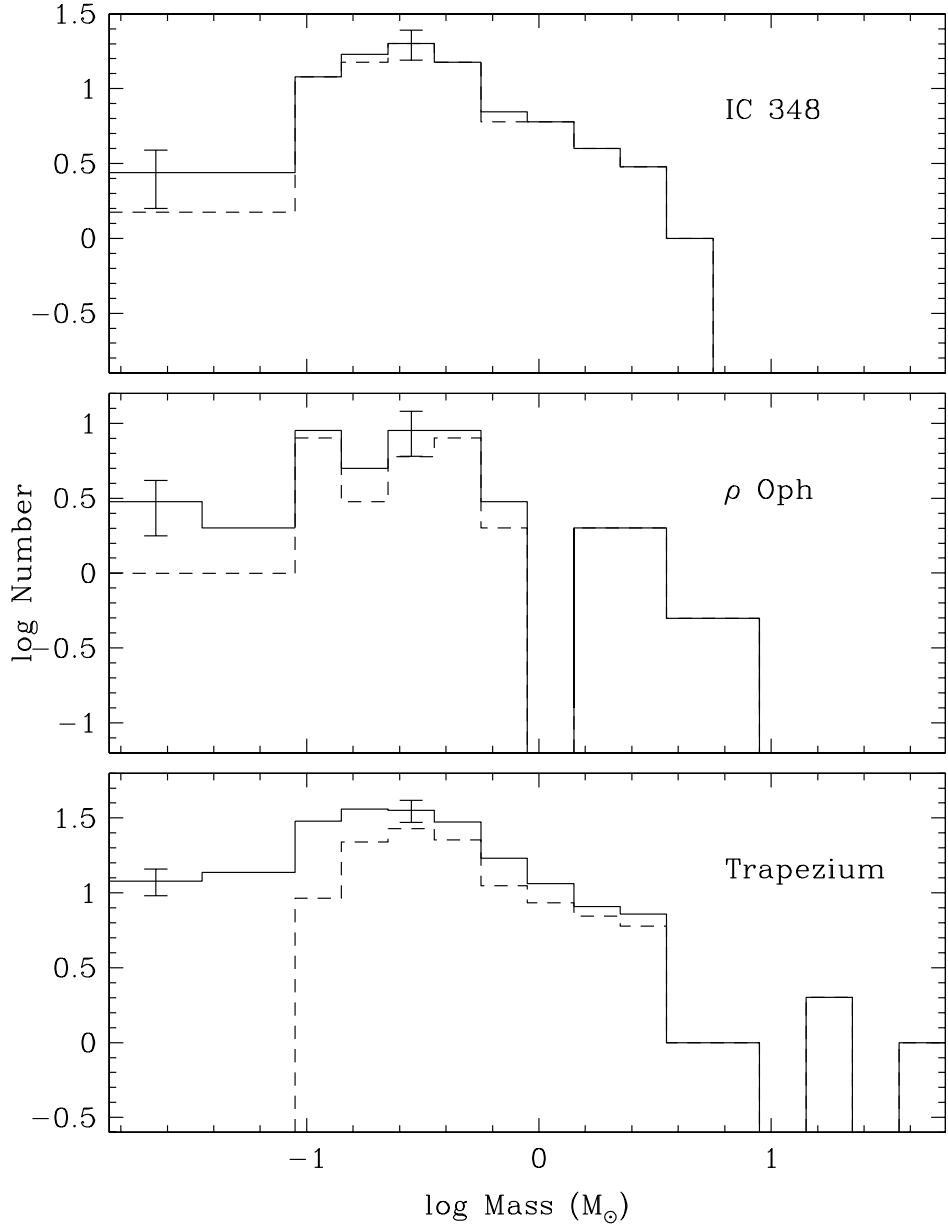}
\caption{
IMFs for the core of IC~348 ($5\arcmin\times5\arcmin$; LRLL updated with
data from Luhman 1999), the cloud core of $\rho$~Oph (LR99), 
and the Trapezium Cluster ($140\arcsec\times140\arcsec$) derived from the 
evolutionary models of DM97. 
The dashed histograms are measured from the spectroscopic samples in each
region and the solid histograms include likely cluster members that lack
spectral types. The two lowest mass bins are given widths of 
$\Delta{\rm log}~M=0.4$ because of the uncertainties in mass estimates. 
The IMFs are complete to log~$M=-1.45$ and $\gtrsim50$\% complete in the bin
from log~$M=-1.45$ to $-1.85$.  To account for the segregation of the OB stars
to the center of the Orion cluster, the values in the highest mass bins 
($>5$~$M_{\odot}$) would be reduced by roughly an order of magnitude.
The counting uncertainties are 
indicated by the error bars in the last bin and at the peak of the IMF. 
}
\label{fig:imf1}
\end{figure}
\clearpage

\begin{figure}
\epsscale{0.85}
\plotone{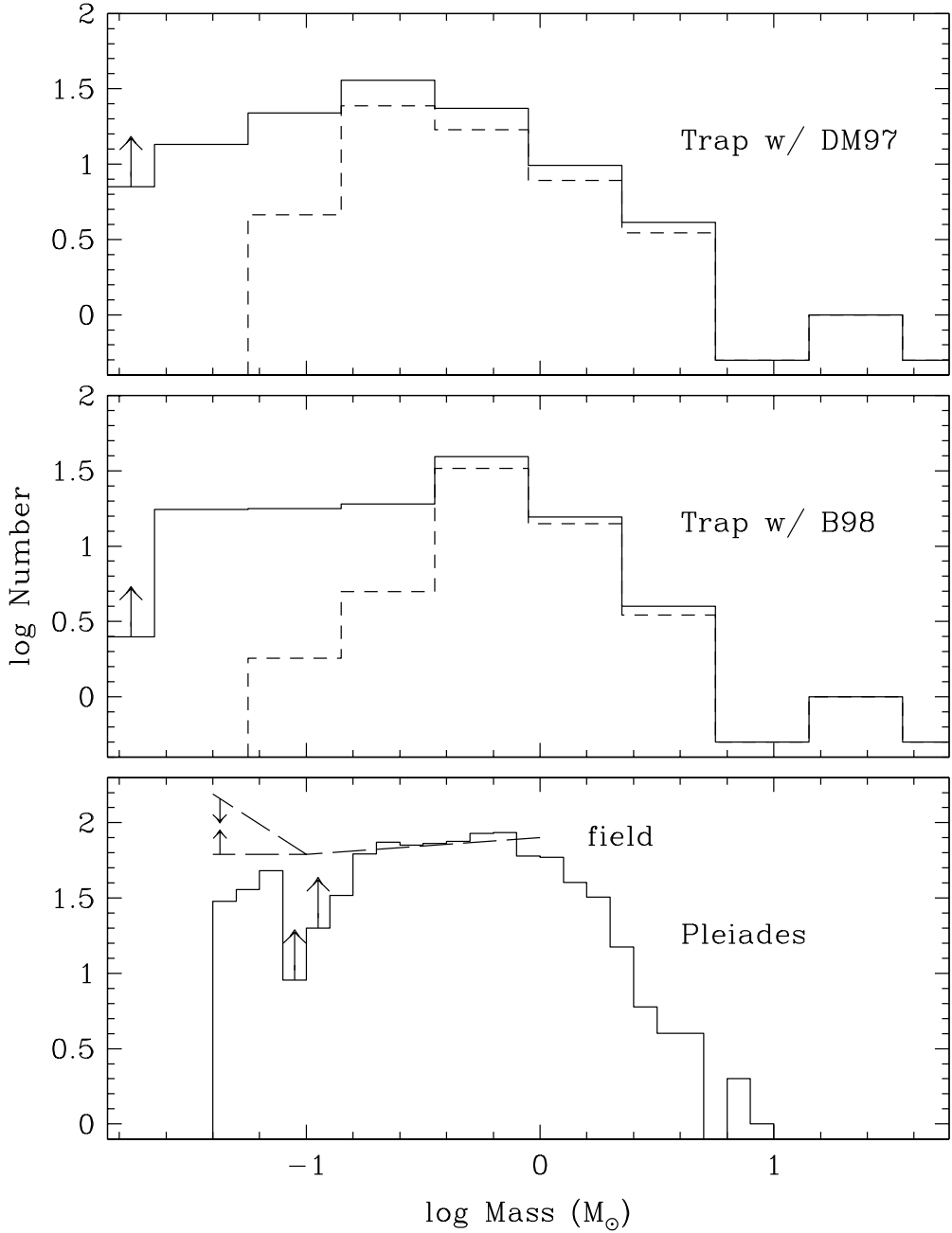}
\caption{
Trapezium IMFs inferred from the evolutionary models of DM97 
({\it upper panel}; same as Figure~\ref{fig:imf1} but with larger bins) and 
B98 ({\it middle panel}). The B98 IMF is our best estimate of the true
form of the IMF in the Trapezium, where the lowest mass bin is a lower 
limit because of incompleteness. The IMFs measured for
the field (Reid et al.\ 1999) and the Pleiades (Bouvier et al.\ 1998)
are given for comparison ({\it lower panel}). As indicated by the 
arrows and two dashed lines, Reid et al.\ (1999) constrained the 
substellar field mass function to have $0\lesssim\alpha\lesssim1$. 
}
\label{fig:imf2}
\end{figure}
\clearpage

\begin{figure}
\epsscale{0.8}
\plotone{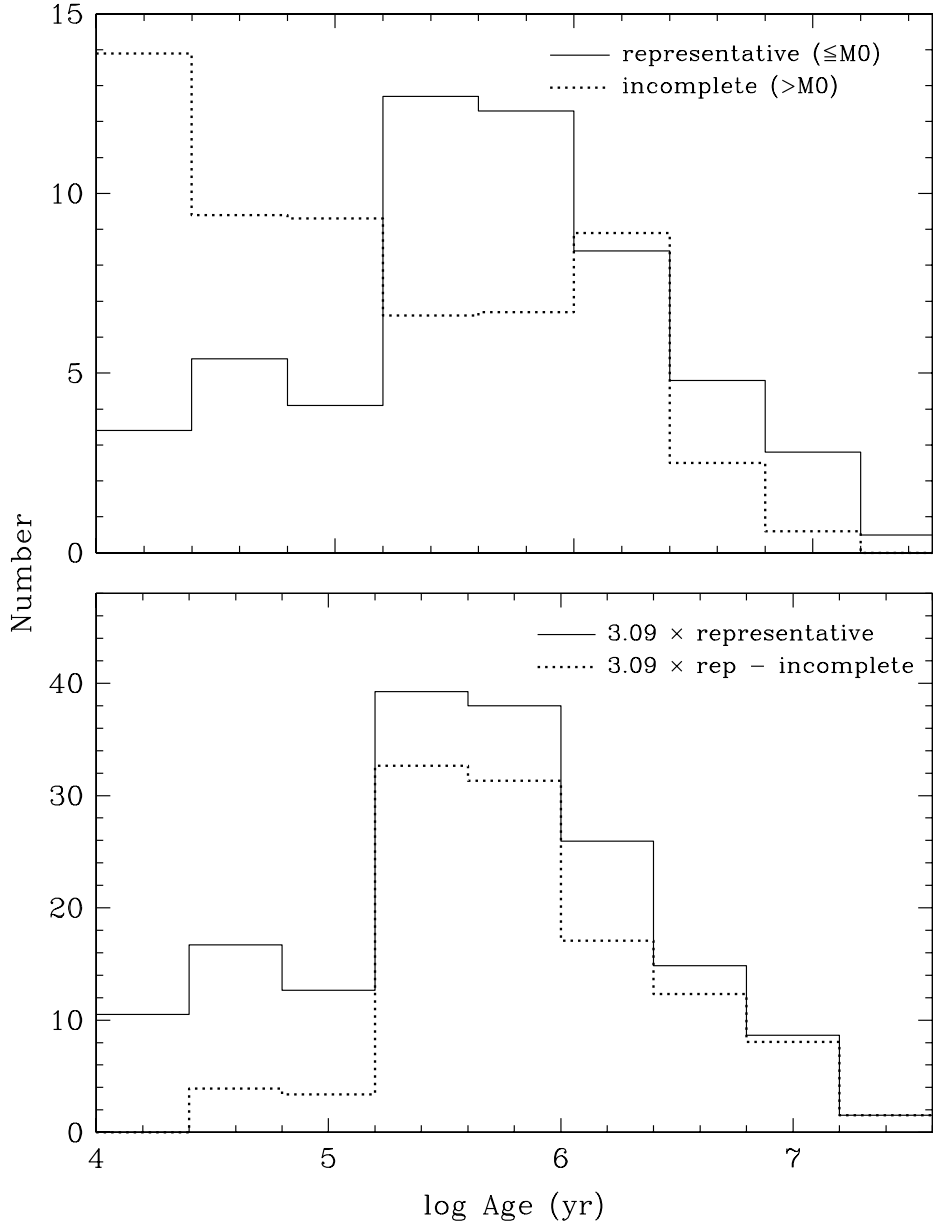}
\caption{
In the top panel, the distribution of DM97 ages for the Trapezium spectroscopic
sample (Figure~\ref{fig:ages1}) is divided into earlier ({\it solid line}) and 
later type sources ({\it dotted line}). The sample with spectral types earlier 
than M0 should reflect the age distribution of the Trapezium, whereas the 
cooler objects are biased towards younger ages because of the spectroscopic 
completeness limit. In the lower panel, the early-type sample's representative 
distribution of ages is normalized ({\it solid line}) to the total number of 
objects that are likely to have types later than M0 (stars classified 
as later than M0 plus all objects 
with $A_H\leq1.4$ lacking spectral types). This distribution minus the 
stars classified later than M0 produces a histogram of ages ({\it dotted line 
in lower panel}) that is used in estimating the masses of the objects lacking 
spectral types. These objects are added as a completeness correction to the IMF 
of the Trapezium in Figure~\ref{fig:imf1}.
}
\label{fig:ages2}
\end{figure}
\clearpage

\begin{figure}
\epsscale{0.85}
\plotone{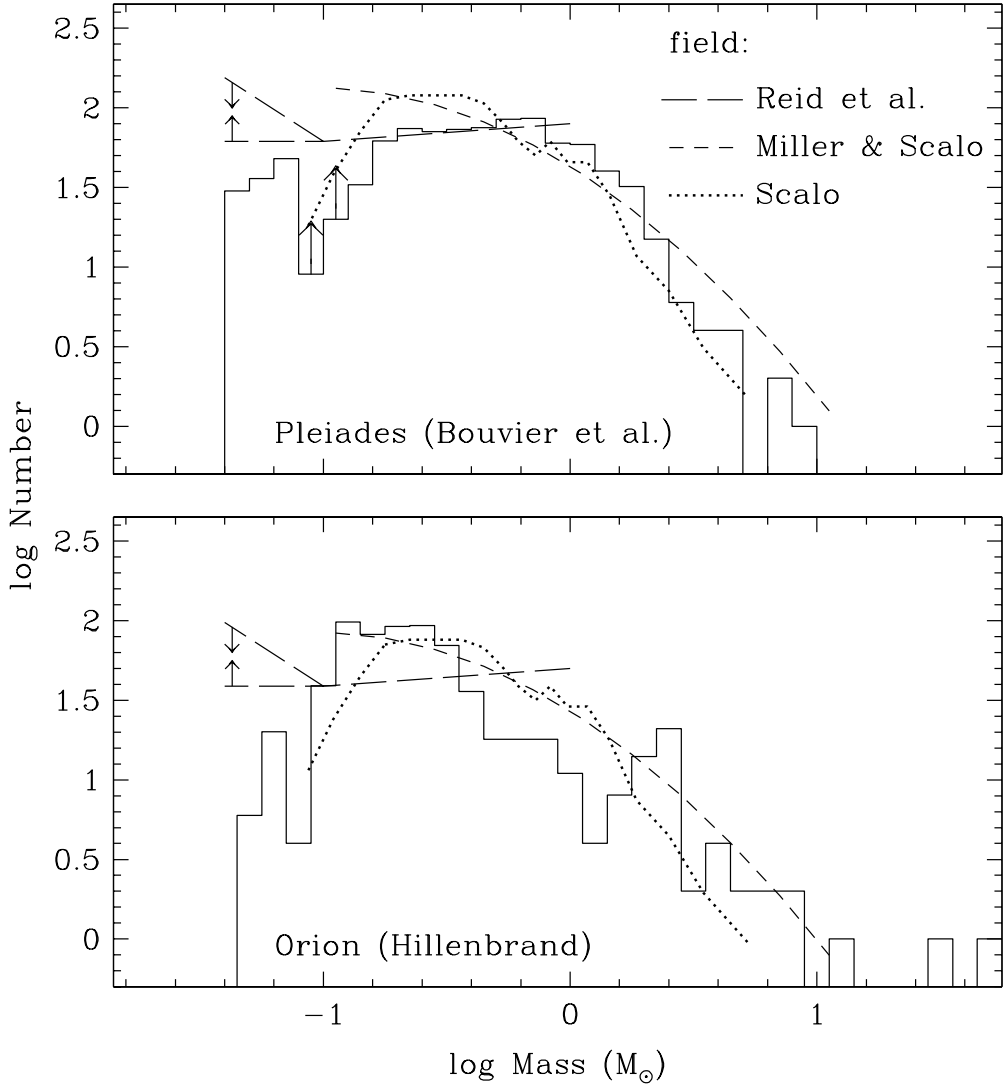}
\caption{
IMFs reported for the Pleiades and Orion by Bouvier et al.\ (1998) 
and Hillenbrand (1997) are compared to the recent field mass function
($<8$~pc) of Reid et al.\ (1999), where the slope of the field substellar mass 
function is constrained to be $0\lesssim\alpha\lesssim1$. The IMF of 
Hillenbrand (1997) is representative for masses above 0.1~$M_{\odot}$.
The mass functions of 
Miller \& Scalo (1979) and Scalo (1986) are also shown for reference. 
}
\label{fig:imfprev}
\end{figure}
\clearpage

\begin{figure}
\epsscale{0.85}
\plotone{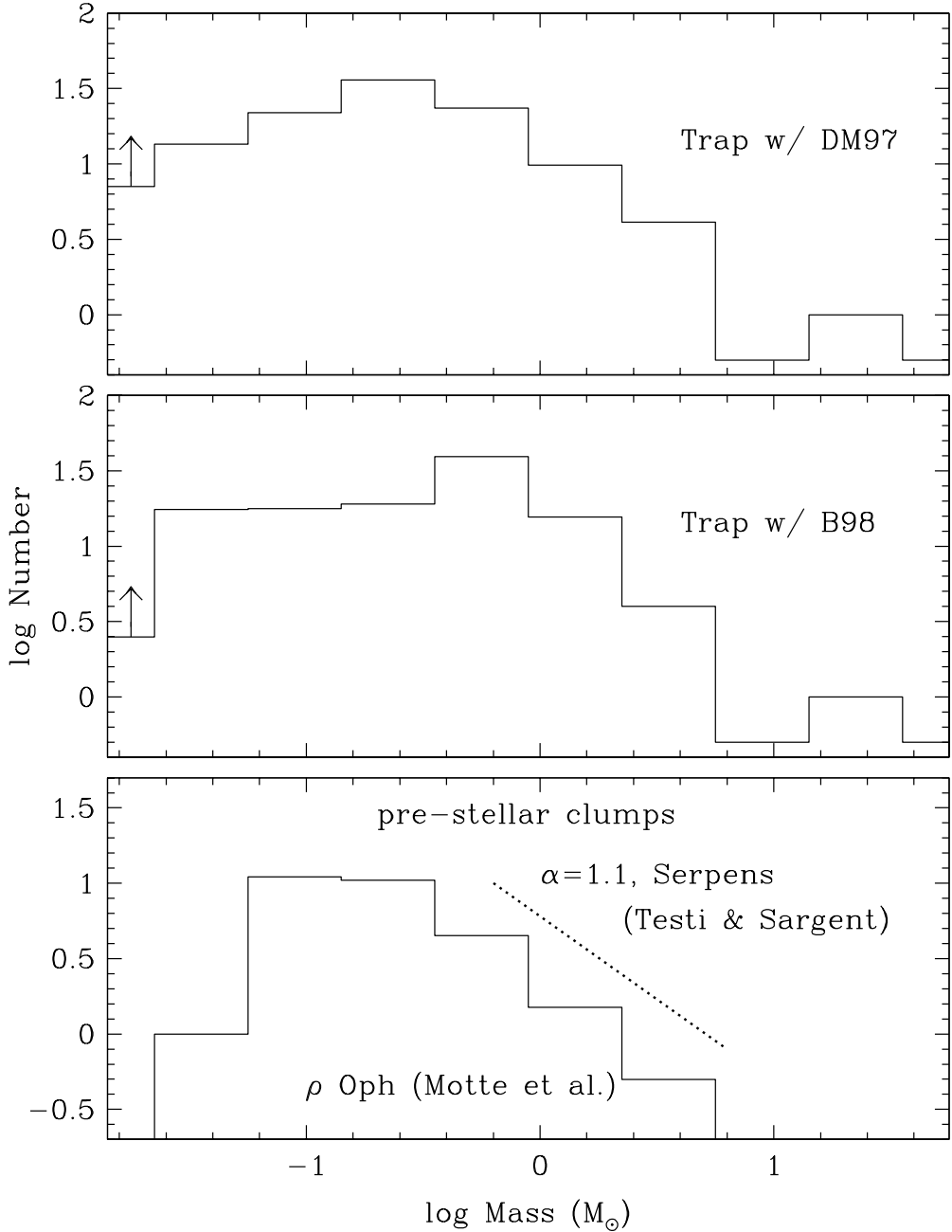}
\caption{
Trapezium IMFs inferred from the evolutionary models of DM97 and B98 are
compared to the mass functions of pre-stellar clumps in Serpens (best fit
power-law; Testi \& Sargent 1998) and $\rho$~Oph (histogram; 
Motte et al.\ 1998). The clump mass functions
are incomplete below a few tenths of a solar mass. 
}
\label{fig:clumps}
\end{figure}
\clearpage

\begin{figure}
\plotone{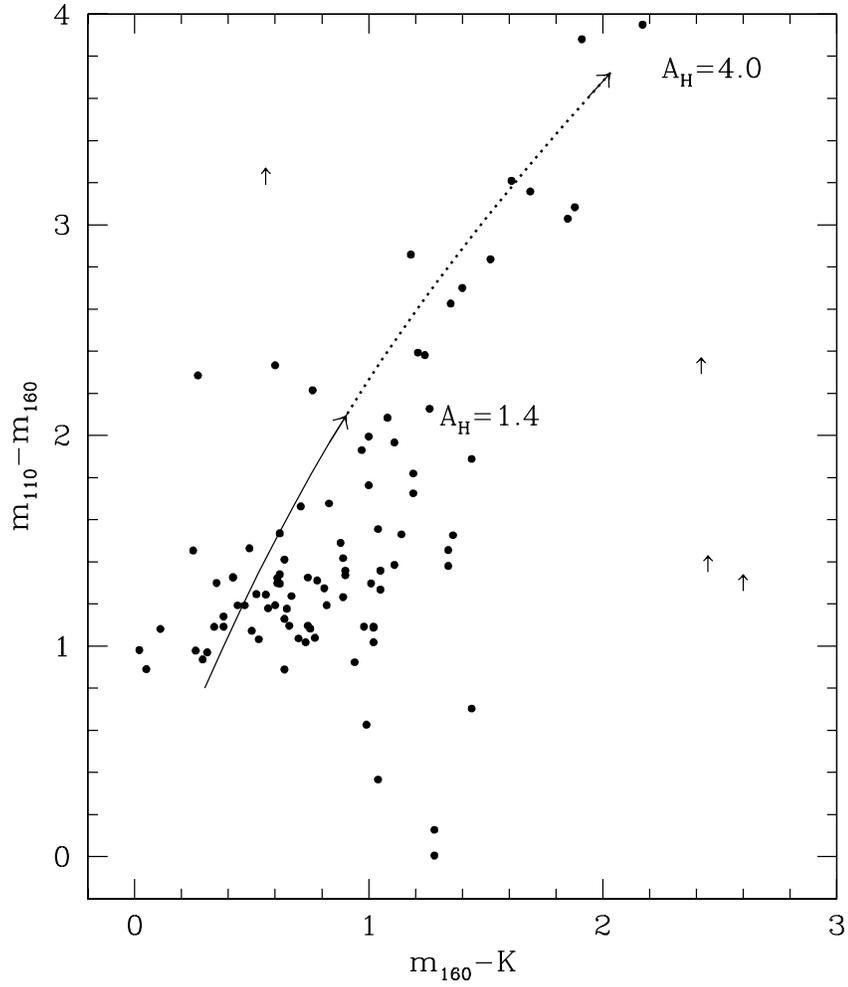}
\caption{
m$_{160}-K$ vs.\ m$_{110}-{\rm m}_{160}$ for stars within $K$-band 
images of the central square arcminute of the Trapezium Cluster (MS). 
Variability between the $K$-band and NICMOS observations is likely for 
some of these young stars, accounting for some of the scatter in m$_{160}-K$.
}
\label{fig:jhhk}
\end{figure}
\clearpage
 
